\documentclass{aa}  
\usepackage{xcolor}
\usepackage{appendix}
\usepackage{graphicx}
\usepackage{txfonts}
\bibpunct{(}{)}{;}{a}{}{,} 

\usepackage{newtxtext,newtxmath}

\usepackage[T1]{fontenc}
\usepackage{ae,aecompl}
\usepackage{natbib}
\usepackage{lscape}

\usepackage{amsmath}	
\usepackage[normalem]{ulem}
\useunder{\uline}{\ul}{}
\usepackage{placeins}
\usepackage{rotating}
\usepackage{booktabs}    
\usepackage{float}
\usepackage{threeparttable}

\newcommand{\kssp}{{$\rm K_s$}\ }
\newcommand{\ks}{{$\rm K_s$}}
\newcommand{\ksnosp}{{$\rm K_s$}}

\begin{document} 
   \title{Unveiling VVV/WISE Mira variables on the far side of the Galactic disk}
   \subtitle{Distances, kinematics and a new extinction law.}

   \author{Rogelio Albarracín
          \inst{1,2}
          \and
          M. Zoccali\inst{1,2}
          \and
          J. Olivares Carvajal\inst{1,2}
          \and
          Á. Rojas-Arriagada\inst{3,2,4,5}
          \and
          J. H. Minniti\inst{6}
          \and
          M. Catelan\inst{1,2}
          \and 
          M. De Leo\inst{1, 2, 7}
          \and
          F. Gran\inst{8}
          \and
          R. Contreras Ramos\inst{1,2}
          \and
          Á. Valenzuela Navarro\inst{1, 2}
          \and
          C. Salvo-Guajardo\inst{1, 2}
          }

   \institute{Instituto de Astrof\'isica, Pontificia Universidad Cat\'olica de Chile, Av. Vicu\~na Mackenna 4860, 782-0436 Macul, Santiago, Chile\\
              \email{rialbarracin@mpia.de}
         \and
             Millennium Institute of Astrophysics, Av. Vicu\~na Mackenna 4860, 82-0436 Macul, Santiago, Chile
         \and
             Departamento de F\'isica, Universidad de Santiago de Chile, Av. Victor Jara 3659, Santiago, Chile
         \and
             N\'ucleo Milenio ERIS
         \and
             Center for Interdisciplinary Research in Astrophysics and Space Exploration (CIRAS), Universidad de Santiago de Chile, Santiago, Chile
         \and
             Department of Physics and Astronomy, Johns Hopkins University, Baltimore, MD 21218, USA
         \and
            Dipartimento di Fisica e Astronomia “Augusto Righi”, Alma Mater Studiorum, Università di Bologna, Via Gobetti 93/2, I-40129, Bologna, Italy
         \and
            Université Côte d’Azur, Observatoire de la Côte d’Azur, CNRS, Laboratoire Lagrange, 06304 Nice, France
             }

   \date{Accepted 05/11/2024. Received 30/08/2024}

\abstract{


\textit{Context}. The structure and kinematics of the Milky Way disk are largely inferred from the solar vicinity. To gain a comprehensive understanding, it’s essential to find reliable tracers in less-explored regions like the bulge and the far side of the disk. Mira variables, which are well-studied and bright standard candles, offer an excellent opportunity to trace intermediate and old populations in these complex regions.\\
\textit{Aims}. We aim to isolate a clean sample of Miras in the Vista Variables in the Vía Láctea survey using Gaussian process algorithms. This sample will be used to study intermediate and old age populations in the Galactic bulge and far disk.\\
\textit{Methods}. Near- and mid-infrared time-series photometry were processed using Gaussian Process algorithms to identify Mira variables and model their light curves. We calibrated selection criteria with a visually inspected sample to create a high-purity sample of Miras, integrating multi-band photometry and kinematic data from proper motions.\\
\textit{Results}. We present a catalog of 3602 Mira variables. By analyzing photometry, we classify them by O-rich or C-rich surface chemistry and derive selective-to-total extinction ratios of $A_{K_{s}}/E(J - K_{s}) = 0.471 \pm 0.01$ and $A_{K_{s}}/E(H - K_{s}) = 1.320 \pm 0.020$. Using the Mira period-age relation, we find evidence supporting the inside-out formation of the Milky Way disk. The distribution of proper motions and distances aligns with the Galactic rotation curve and disk kinematics. We extend the rotation curve up to R$_{\rm GC} \sim 17 \ \rm{kpc}$ and find no strong evidence of the nuclear stellar disk in our Mira sample. This study constitutes the largest catalog of variable stars on the far side of the Galactic disk to date.}


   \maketitle



\section{Introduction}

Although we are able to study the Milky Way (MW) galaxy in great detail, de-projecting the position of the stars in the sky into a three-dimensional map is still challenging due to the difficulty of deriving precise distances for individual stars. This task is especially hard for stars in the Galactic plane, as dust extinction, heavily affecting distances and often degenerated with it, is significantly larger. Given that reliable geometric distances are currently available only within a few kpc from the Sun \citep[]{bailer-jones+23}, we rely on specific stellar tracers, for which we are able to derive distances because we have independent information about their intrinsic luminosity. 

Among tracers, pulsating variable stars are the most common ones, as most of them follow tight period-luminosity (P-L) relations \citep{Marciobook}. RR Lyrae stars have been used to trace the structure of the old component of the Galactic bulge \citep[e.g.,][]{Dekany2013,pietrukowicz+15,Prudil+19,Du+2020,JOlivares+24}, and Classical Cepheids have proven to trace the structure of the near and far Galactic disk \citep[e.g.,][]{skowron19, Minniti+2021}. The nature of these classes of variable stars restricts the analysis to strictly old or young stellar populations, respectively. Mira variables, in turn, offer a unique possibility to trace the range of intermediate-age populations in our Galaxy.

Mira variables are highly-evolved stars from the thermally pulsating asymptotic giant branch (AGB). AGB stars exhibit pulsation in different modes, with complex interactions between the pulsation, the convective envelope, and the stellar atmosphere \citep{Freytag+2017, Hofner+2018, Chiavassa+2024}. The modes and regularity of the pulsation between cycles define different sub-classes of long-period variables (LPVs) occupying different P-L sequences \citep{Wood2015}. Indeed, observations of the LMC have shown that  LPVs lie in distinct sequences depending on their sub-class \citep{Soszynski+2009}. Mira variables, in particular, pulsate in the fundamental mode, exhibiting amplitudes of $\Delta V > 2.5 \ \mathrm{mag}$ with periods ranging from 80 days to over 1000 days, following a single P-L sequence \citep{Glass82}. Mira variables have been proven to follow tight P-L relations in the near-infrared (near-IR) and infrared bands \citep[e.g.][]{Feast+89, Whitelock+2008, Yuan+2018, Iwanek+2021-PL, SandersPL2023}. They are well suited as standard candles for cosmological applications \citep{Huang2018, Huang+2024}, local group studies \citep{Menzies+2015} and the MW \citep{Matsunaga+09, GBE2020, Urago+2020, Iwanek+23}. A recent study by \citet[][hereafter S22]{Sanders2022} identified a large number of Miras in the MW nuclear stellar disk (NSD) from VISTA Variables in the V\'\i a L\'actea (VVV) data. The existence of Miras in this poorly known component is especially important as their age, constrained through their period, allows for the formation of the main Galactic bar to be dated \citep{BabaKAwata20, sa-freitas+23}. In addition, thanks to their precise distances and proper motion, they allow for models of the NSD to be computed \citep{Sanders+2024}.  


\begin{figure*}
\includegraphics[width=\hsize]{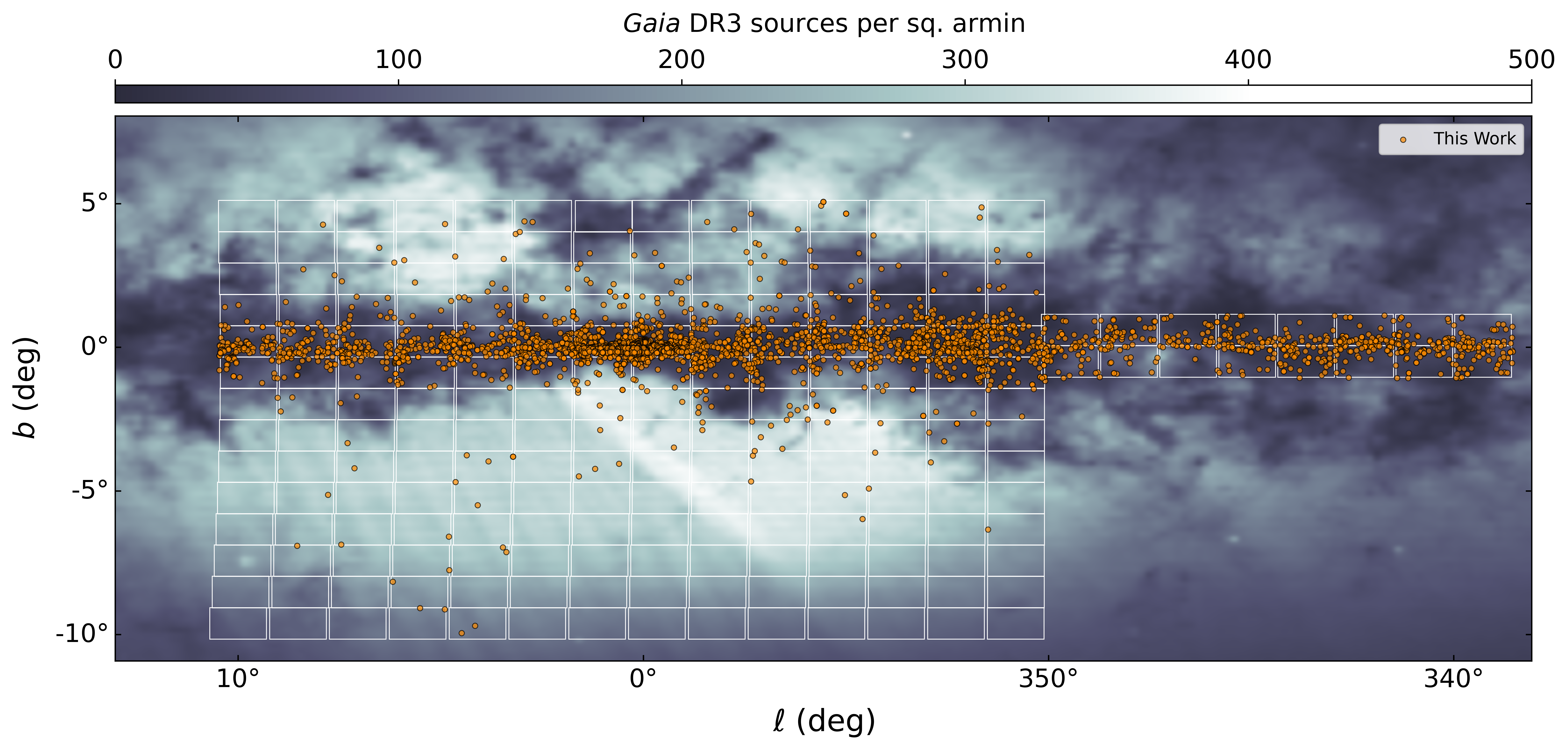}
    \caption{Gaia DR3 source density around the region covered by the VVV survey. White boxes represent VVV tiles considered in this study. Orange dots represent Mira variables identified by our GP algorithm. Color indicate the amount of sources in the {\textit Gaia} DR3 catalog. Density map data were taken from the Gaia archive.}
    \label{fig:sky}
\end{figure*}

At a given mass, any thermally pulsating AGB star begins to show fundamental-mode pulsations at a very narrow range of radii \citep{Trabucchi+2019}. Thus, a dependence of the pulsation period with stellar mass, and therefore age, is expected. Recent theoretical studies of this relation can be found in the literature \citep[see][]{eggen98, TrabMowlavi2022}. However, most period-age relations have been calibrated empirically. Stars near the Sun with hotter kinematics are, on average, older \citep{DeSimone+2004}, and indeed, this relation has been long observed for Miras \citep{Merrill1923, Feast63}. Using the exquisite astrometric precision of {\it Gaia} \citep{gaia_mission}, \citet{Zhang+2023} have obtained a period-age (P-A) relation, using the velocity dispersion of Mira variables in the solar neighborhood. As discussed in both \citet{Nikzat+2022} and \citet{Zhang+2023}, the exact shape of this relation is still highly debated, and some key anchors, such as LMC clusters, can have significant contamination. Although the correlation between period and age is not under discussion, a spread in the relation is expected. \citet{TrabMowlavi2022} used a set of theoretical pulsation models to show that a Mira with a period of 350 days can have ages that differ up to 3 Gyr. 

Low and intermediate-mass stars in the AGB also undergo significant mass loss \citep{Hofner+2018}. Strong convection and shockwaves due to pulsation generate winds that lead to cool envelopes, where dust and molecule formation are expected \citep{Freytag+2023}. Changes in the surface ratio of C/O, due to the third dredge-up, affect the dominant molecular species formed in the envelope, from oxygen-based molecules such as TiO and SiO to carbon-based molecules like CN and $\mathrm{C}_{2}$ \citep{Cristallo+2011}. This transition in chemical abundance greatly impacts the spectral energy distribution \citep{Iwanek+23}, and, in turn, the mean magnitudes and the corresponding P-L relation \citep{Whitelock+03,Soszynski+2005,Yuan+17, Iwanek+2021-PL}. Therefore, an assessment of the Mira chemical class is crucial for adequate distance determination.

A challenge in the characterization of light curves from Miras is their long-term trends and period variations. Indeed, it has been observed that Mira variables show long-term modulations independent from the main pulsation period \citep[e.g.][]{Zijlstra+2002, Templeton+2005, Uttenthaler+2011, Ou+2022}. Multiple mechanisms responsible for these trends have been proposed; stellar winds triggered by thermal pulses \citep{Vassiliadis93}, coupling of the pulsation with convection \citep{Freytag+2017} or presence of dust shells \citep{Zijlstra+96, Ou+2022}. A work frame that accommodates these variations in the modeling is necessary for adequate identification of Miras.




In this paper, we aim to identify Mira variables in the VVV survey by means of a Gaussian Process (GP) algorithm to study the 3D distribution of these stars along with the kinematics via proper motions (PMs). In Section \ref{sec:Data}, we describe the data considered from different surveys, and Section \ref{sec:Identification} describes the algorithm used for the identification of Miras and modeling of the light curves. Section \ref{sec:SampleProp} explains the separation of C-rich and O-rich stars in our sample and the determination of an extinction law. Section \ref{sec:3Ddistrib} presents the 3D spatial distribution of the stars. Section \ref{sec:kinematics} presents the kinematics of the sample. Finally, in Section \ref{sec:Discussion_Conclusions}, we present the discussion and conclusions of our study.



\section{Observations and photometry}
\label{sec:Data}
\subsection{VVV data}

The catalog of Mira variables presented in this work is based on Point Spread Function (PSF) fitting photometry performed on multi-epoch near-IR data from the VVV ESO public survey \citep{minniti10}. Since 2010, the VVV survey has observed nearly 1700 deg$^2$ over the Milky Way bulge and southern disk. Further data is available as a part of VVV eXtension \citep[VVVx][]{Minniti2018}, extending the baseline to roughly 13 years. In this study, we use the full footprint of the bulge region, from $-10^{\circ}$$\lesssim$$\ell$$\lesssim$+$10^{\circ}$ and $-10^{\circ}$$\lesssim b \lesssim$+$5^{\circ}$, together with tiles from the disk, from $-22^{\circ}$$\lesssim$$\ell$$\lesssim$$-10^{\circ}$ and $-$1$\lesssim b \lesssim$1, encompassing tiles b201 to b396, d069 to d076 and d107 to d114. The footprint of the catalog in the sky can be seen in Fig.\ref{fig:sky}. PSF-based photometry was carried out across all available epochs, including $\sim$100 in the \kssp band and fewer than 12 in each of the J and H bands. The average seeing is $\sim$1 arcsec, as images with seeing larger than 1.7 arcsec were discarded from the analysis.

Multi-epoch photometry was derived using a photometric pipeline based on the DAOPHOT II/ALLSTAR package \citep{stetson+87}. An initial, high-threshold photometry was performed to detect bright stars and coordinate transformation across all available epochs in each VIRCAM chip. The images were stacked accordingly, and deep, low-threshold photometry was then performed, creating the most complete stellar master list. Using this master list, PSF photometry was performed on each epoch, allowing the centroids for each star to be refined. This resulted in a list of stars with consistent identification across all epochs and a position that can change in time. Instrumental magnitudes were normalized to the photometric system of a reference epoch by means of the DAOMATCH/DAOMASTER codes \citep{stetson+87}. Only this reference epoch was then photometrically calibrated to the 2MASS reference system, as described below. Proper motions, on the other hand, were derived as described in \citet{contrerasramos17}, following the prescriptions from \cite{Anderson2006} and \cite{Bellini2014}. 

The photometry has been calibrated to the VVV photometric system as explained in detail in \citet{zoccali+24}. Briefly: bright, isolated stars in common with the 2MASS catalog \citep{2mass} were identified within each VIRCAM detector. The 2MASS magnitudes were transformed to the VVV photometric system using the color and extinction terms reported in \citet{gf+18_VVVcal}. A zero point was then derived in each band as the median difference between the VVV instrumental magnitudes and the transformed 2MASS magnitudes and applied to the former.

By construction, the PM of each individual star was derived with respect to the average motion of a local sample of bulge Red Giant Branch (RGB) stars. Therefore, they are relative PMs. Nonetheless, they were later calibrated to the {\it Gaia} data release 3 (DR3) astrometric system by means of $>$80 common stars, selected within a quarter of a VVV detector. This is the unit area for which we derive a common astrometric calibration to the {\it Gaia} system \citep[see][for more details about this step]{zoccali+24}.

\subsection{WISE data}

In addition to the VVV photometry, we complemented the variability information with data from the Wide-field Infrared Survey Explorer \citep[WISE][]{WISEcat}. WISE is a 40 cm space telescope with the mission of mapping the entire sky in four mid-IR bands; W1 ($\lambda_{\rm{eff}} = 3.4 \mu m$), W2 ($\lambda_{\rm{eff}} = 4.6 \mu m$), W3 ($\lambda_{\rm{eff}} = 12 \mu m$) and W4 ($\lambda_{\rm{eff}} = 22 \mu m$). The initial WISE mission was conducted from 2010 to 2011, after which the telescope was reactivated as part of the Near-Earth Object WISE Reactivation Mission \citep[NEOWISE-R;][]{neowiser2011}, providing additional data in the W1 and W2 bands from 2013 to the present. We cross-matched all sources from the VVV catalog with the WISE Multiepoch Photometry Table \citep{WISE_multiepoch}\footnote{https://wise2.ipac.caltech.edu/docs/release/allwise/} and the NEOWISE-R Single Exposure (L1b) Source Table \citep{Neowise_single} using the NASA/IPAC Infrared Science Archive \footnote{https://irsa.ipac.caltech.edu/applications/Gator/}. Matches were performed with a tolerance of 1 arcsecond, and magnitudes were corrected for saturation effects following the prescriptions in \citet{WISEcorr}. Out of our final sample of 3,602 stars, 2782 have measurements in the W1 and W2 bands, 2242 in W3, and 2027 in W4. The median and maximum values for individual data points in each of the WISE bands were 23, 22, 11, 11, and 35, 32, 14, 14 for W1, W2, W3, and W4, respectively. We note that even though the magnitudes are corrected for saturation, several matched sources fall within the saturation range for the WISE bands, specifically 593 in W1, 752 in W2, 382 in W3, and 24 in W4.

Furthermore, the large PSF of WISE is likely to cause blending in highly crowded regions near the Galactic plane. Since Mira variables are very bright in the mid-infrared, we consider the effects of contamination to be minor for the mid-infrared color excess used in Sect. \ref{sec:OC_composition}. However, for precise distance determinations, the contribution of blended sources in the WISE magnitudes affects both the measured median magnitude and the extinction estimates. We expand on this issue in Sect. \ref{sec:distances} and Appendix \ref{sec:WISE_distances}.

\subsection{Variable Star Zoo Mira Candidates}

In 2018, our group started a citizen science project named Variable Star Zoo, which was hosted at the Zooniverse platform\footnote{\url{https://www.zooniverse.org}}. The users were asked to classify variable stars by comparing their phased light curves with some predefined templates. An input catalog contained roughly 55,000 candidate variable stars from the VVV survey, located within $-10^{\circ}$$\lesssim$$\ell$$\lesssim +10^{\circ}$ and $-1.6^{\circ}$$\lesssim b \lesssim 1.8^{\circ}$. The available templates included two eclipsing binaries, an RR-Lyrae and a Cepheid, a Mira, a microlensing event, as well as an unusual variable source and a pure-noise light curve. A candidate variable was considered classified, hence removed from the input catalog as soon as it reached 15 classifications. A more extensive discussion of the Variable Star Zoo Project is presented in \citet{Huanca2022}. For this work, we initially selected the $\sim$2000 stars that were classified as Mira by 11 or more volunteers. Subsequent visual inspection carried out independently by four team members reduced the sample to 1883 Mira candidates, which, hereafter, will be referred to as the VarZoo sample. By using the VarZoo sample as a calibrator for the algorithms explained in Sect. \ref{sec:Identification}, we constructed our final Mira sample, shown in the sky in Fig.~\ref{fig:sky}.


\section{Mira variable identification}

\label{sec:Identification}

\begin{figure}
\includegraphics[width=\hsize]{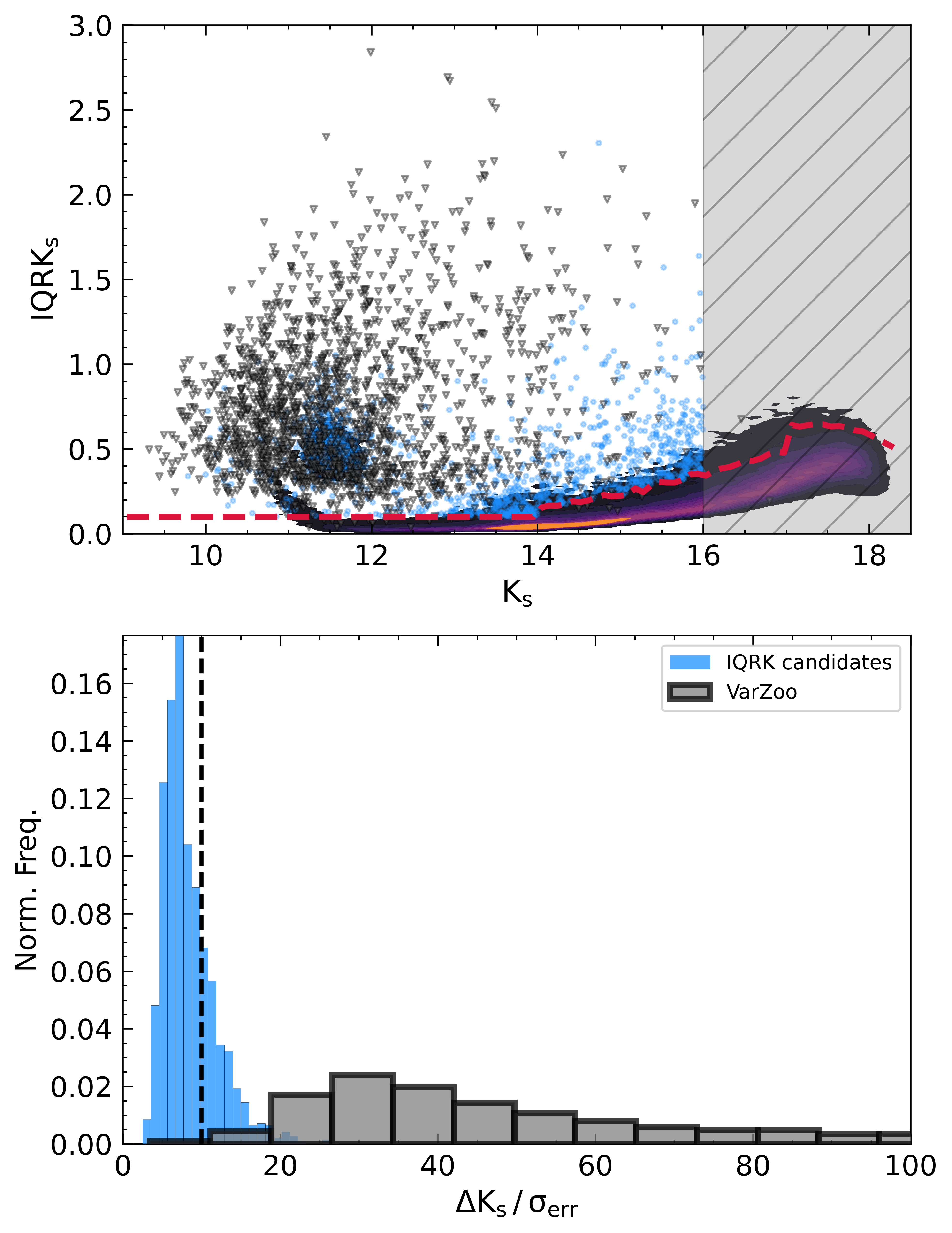}
    \caption{Amplitude-based cuts in Inter Quartile Range in \kssp band and Amplitude over error for an example VVVx tile. \textit{Top}: Blue points show candidates that meet IQRKs and amplitude over noise criteria. The red dashed line shows the IQRKs threshold as a function of \kssp magnitude; black triangles are the control VarZoo sample. Points inside the gray-shaded region are excluded from further analysis. \textit{Bottom}: Histogram of the amplitude of variability over the source-averaged photometric error. Blue histograms are sources that passed the IQRKs cuts, and the gray histogram is the VarZoo sample. The black dotted line defines the threshold value to further consider candidates ($\Delta \rm{K_{s}}/\sigma \geq 10$).}
    \label{fig:IQRKs}
\end{figure}

In order to select candidate Miras, we followed the prescription by S22. We constructed cuts based on an interquartile range of magnitudes in the \kssp band (IQRKs; Fig.~\ref{fig:IQRKs}), which we optimized until all the stars in the parameter space occupied by the VarZoo sample were selected, but not many more. We applied different cuts for brighter and fainter stars in order to separate intrinsic variability from statistical fluctuations due to low signal-to-noise ratio (SNR). For stars brighter than \ks=14 we considered as candidate Mira any star with IQRKs$>$0.1. For sources in the range 14$\leq$\ks$\leq$16 we required IQRKs$\geq$1$\sigma$ above the median value. Sources than satisfy these amplitude cuts are shown as blue dots in the top panel of Fig. \ref{fig:IQRKs}. We neglected sources with \ks$\geq$16 as their light curves would be dominated by the noise (gray shaded area in Fig.~\ref{fig:IQRKs}, top). We are aware that this first cut does not remove spurious variability due to saturation. However, removing sources in the saturation range would remove many real Mira variables. The top panel of Fig.~\ref{fig:IQRKs} shows these cuts applied to a VVVx tile, along with the VarZoo sample.  We see that the majority of VarZoo stars lie above the imposed threshold. To further clean the candidates from artificial variability due to noise, we impose a cut in the ratio of light curve amplitude to the average photometric noise of $\Delta$\ksnosp$/\sigma \geq 10$. From the bottom panel of Fig. \ref{fig:IQRKs}, we see that almost the entire VarZoo sample, represented as a black histogram, lies above the threshold (black dotted line in Fig. \ref{fig:IQRKs}, bottom). We use these cuts to feed our algorithm with an initial list of candidates showing a strong intrinsic variability based on amplitude only. For further consideration as candidates, a strong periodic signal consistent with Mira variables needs to be detected.

Mira variables are characterized by a prominent main oscillation due to pulsation, with periods over 80 days and a \kssp amplitude $\Delta \rm{K_{s}} \gtrsim$ 0.4 mag \citep{Whitelock2006, Iwanek2021}. Together with the main pulsation, Mira and other red giants often show additional variability \citep[see][]{He2016, Matsunaga2017}. The timescale for this additional variability can be of the same order as the main period or up to several times larger \citep{Nicholls2009}. For these reasons, flexible modeling that separates Mira variables from other variability classes and noise while accounting for quasi-periodic variability is necessary.

\begin{figure}[ht]
\includegraphics[width=\hsize]{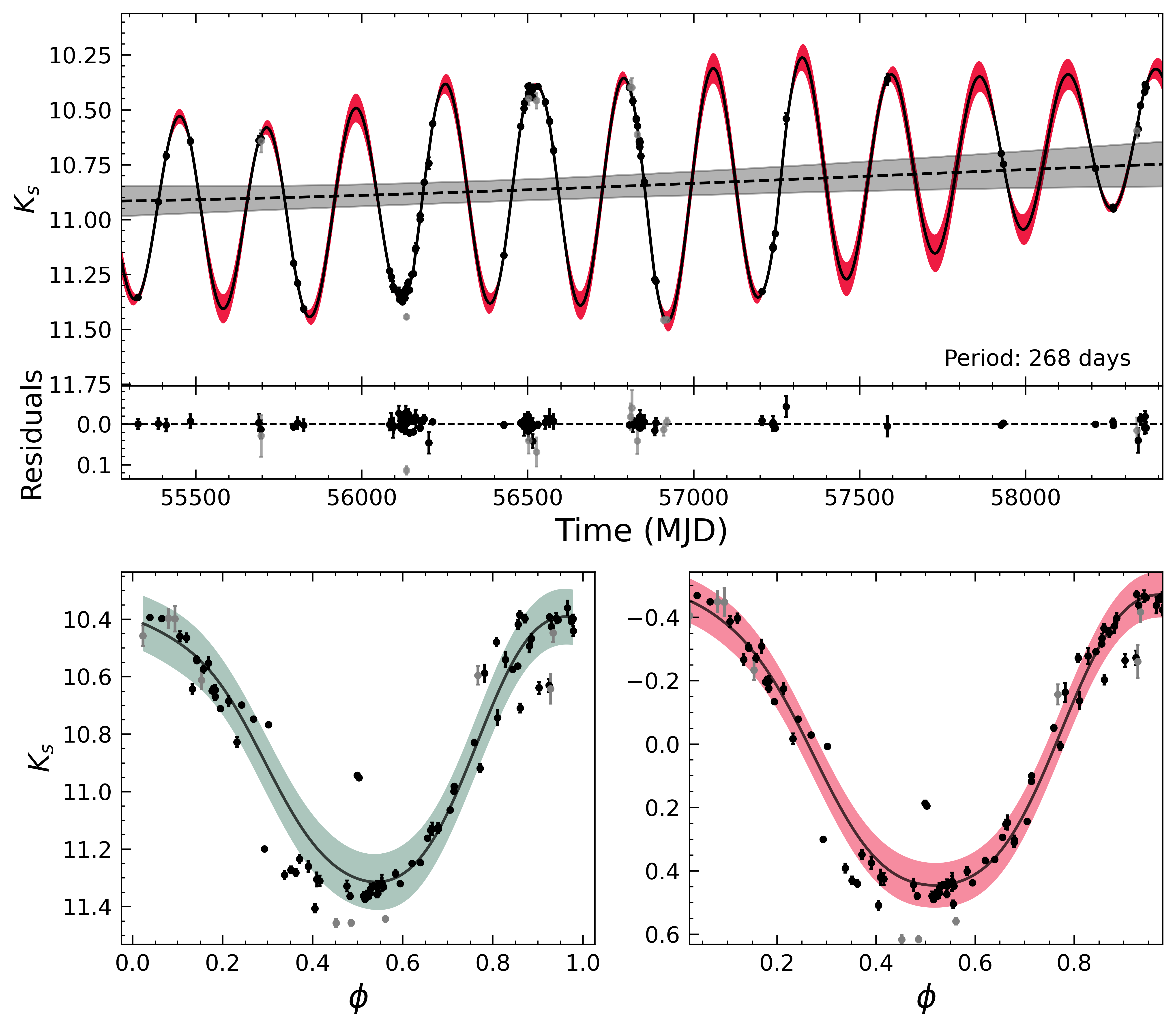}
    \caption{Example of the GP modeling for a single source. \textit{Top}: Time series photometry for this star; red shades denote the confidence intervals for the GP model, black shaded line denotes the median trend with corresponding confidence interval in gray. Gray and black points are masked and accepted points, respectively. \textit{Bottom panels}: Phased light curves for the star without (left) and with (right) the 
    median trend correction applied. Colored regions are the confidence intervals for the Fourier fit.}
    \label{fig:iterative_GP}
\end{figure}

We perform a Fast Lomb-Scargle \citep{Lomb76, Scargle82, PressRybucki89} periodogram\footnote{See \url{docs.astropy.org/en/stable/timeseries/}} on the candidate light curves, retaining sources where any of the best 3 periods, with False Alarm Probabilities (FAPs) less than 10\%, are over 50 days. To identify LPVs, we further inspect the periodicity of the candidates by means of a Generalized Lomb-Scargle (GLS) periodogram \citep{Zechmeister2009}. This method has been reported as having the highest accuracy in recovering Mira periods \citep{Graham2013}. We cleaned the periodogram from period aliases related to the survey sampling rate by using the window method described in \citet{Kramer2023}. 
In some cases, the periodogram powers for the 3 best GLS periods were approximately the same. In such cases
we selected the period with minimal Conditional Entropy as defined by \citet{GrahamCE}.

After rejecting individual data points with errors 3$\sigma$ above the median error, we fed the \kssp band light curves and the selected period to a grid of iterative GP models. Similarly to S22, we used a combination of quasi-periodic and non-periodic kernels in order to capture the behavior of Mira variable light curves, detailed in Eq. (\ref{eqn:kernel}). For each iteration, we selected the model that minimized the Aikake information criterion (AIC). We fitted a simple polynomial to the GP model in order to identify the median trend and subtract it. We phase-folded the de-trended light curve using the period re-adjusted by the GP algorithm. We fitted a Fourier series of three harmonic terms and calculated the distance between each point and the model adequately normalized, as prescribed in (Baeza-Villagra et al. in prep.). Points with a distance over 3$\sigma$ above the median were rejected and not considered for further analysis. The remaining points were fed to the next iteration of the GP, where a new model fitted. For the model of the next iteration to be accepted, the reported likelihood had to exceed that of the previous iteration (i.e., $ \log_{10}{\mathcal{L}}_{i} \leq  \log_{10}{\mathcal{L}}_{i + 1} + 10$), with a maximum of 5 iterations. This procedure allows for flexible modeling of the light curve while accounting for stochastic behavior and saturation effects that otherwise would hamper the performance of the $\mathcal{L}$ parameter as a quality indicator for Mira candidates. An example of the described modeling for a VarZoo Mira candidate is shown in Fig. \ref{fig:iterative_GP}.

\begin{figure*}
\includegraphics[width=\hsize]{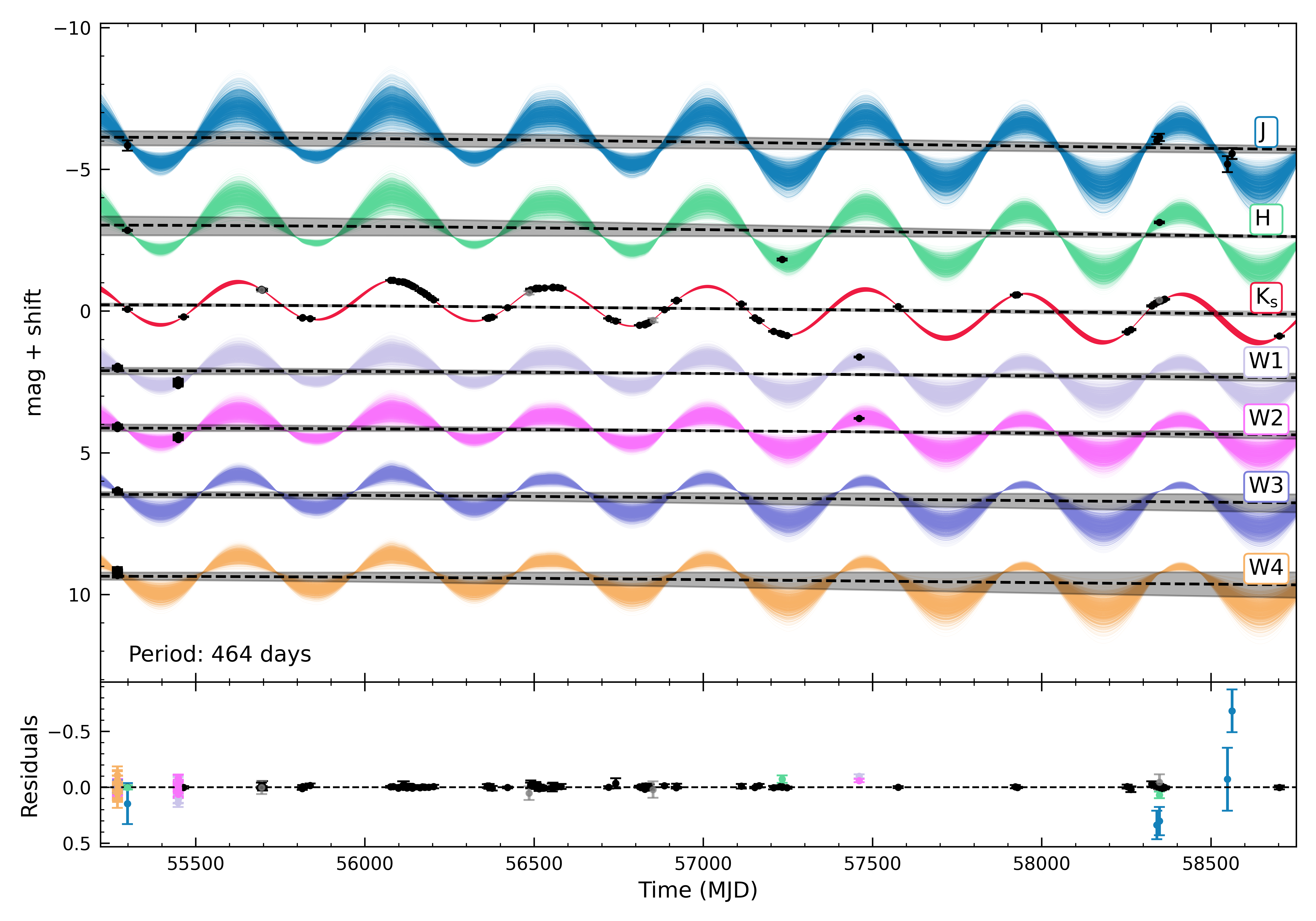}
    \caption{Example multi-band light curve for a source in our catalog. \textit{Top:} Fitted models for a candidate in the different bands, ordered in increasing wavelength from top to bottom: J, H, \ksnosp, W1, W2, W3, W4. Black points are photometric measurements in each band. Solid lines indicate each of the sampled models. Gray dashed lines are the modeled median trends with the respective confidence intervals in black-shaded regions. \textit{Bottom:} Residuals of the band modeling. Points are color-coded according to the observing band, as in the top panel.}
    \label{fig:multiband_GP}
\end{figure*}

\subsection{Multi-band modeling}
\label{sec:Multiband}

The method described above only applies to the \kssp light curves. In the other bands, the number of epochs is, at most, a handful.  Therefore, in order to accurately derive the median magnitudes for the other photometric bands, we constructed a light curve template based on the \kssp data. Specifically, we apply a Fourier transform to the model for the \kssp data obtained in the previous step. To apply this model to other bands, we multiply the Fourier coefficients of the \kssp band by the amplitude ratios derived in S22, which describe how the amplitude in a given band relates to the measured amplitude in the \kssp band. After applying an inverse Fourier transform to the coefficients, a simple fit on the available data in the given band is performed. In order to derive a median magnitude and its uncertainty, we obtained model fit sampling over the confidence intervals of amplitude ratios and the photometric errors. For each combination of sample points and amplitude ratios, we obtained a best-fit model and a corresponding polynomial fit and repeated this procedure 1000 times.
The average of the fitted polynomials is a distribution from which a final median magnitude and uncertainty were obtained. We emphasize that the number of observations per band and their phase have a large impact on the uncertainty of the median magnitude. Although phase shifts are expected in Mira light curves between different bands \citep{Iwanek2021}, we chose not to include these additional parameters due to the small number of data points in bands different from \ksnosp. This phase shift is close to 0 in the near-IR bands and $\Delta \phi \sim 0.1$ for WISE filters \citep{Iwanek2021}. Figure~\ref{fig:multiband_GP} shows an example of the multi-band fit for a source in the VarZoo sample.

\subsection{Candidate selection}

\begin{figure*}
\includegraphics[width=\hsize]{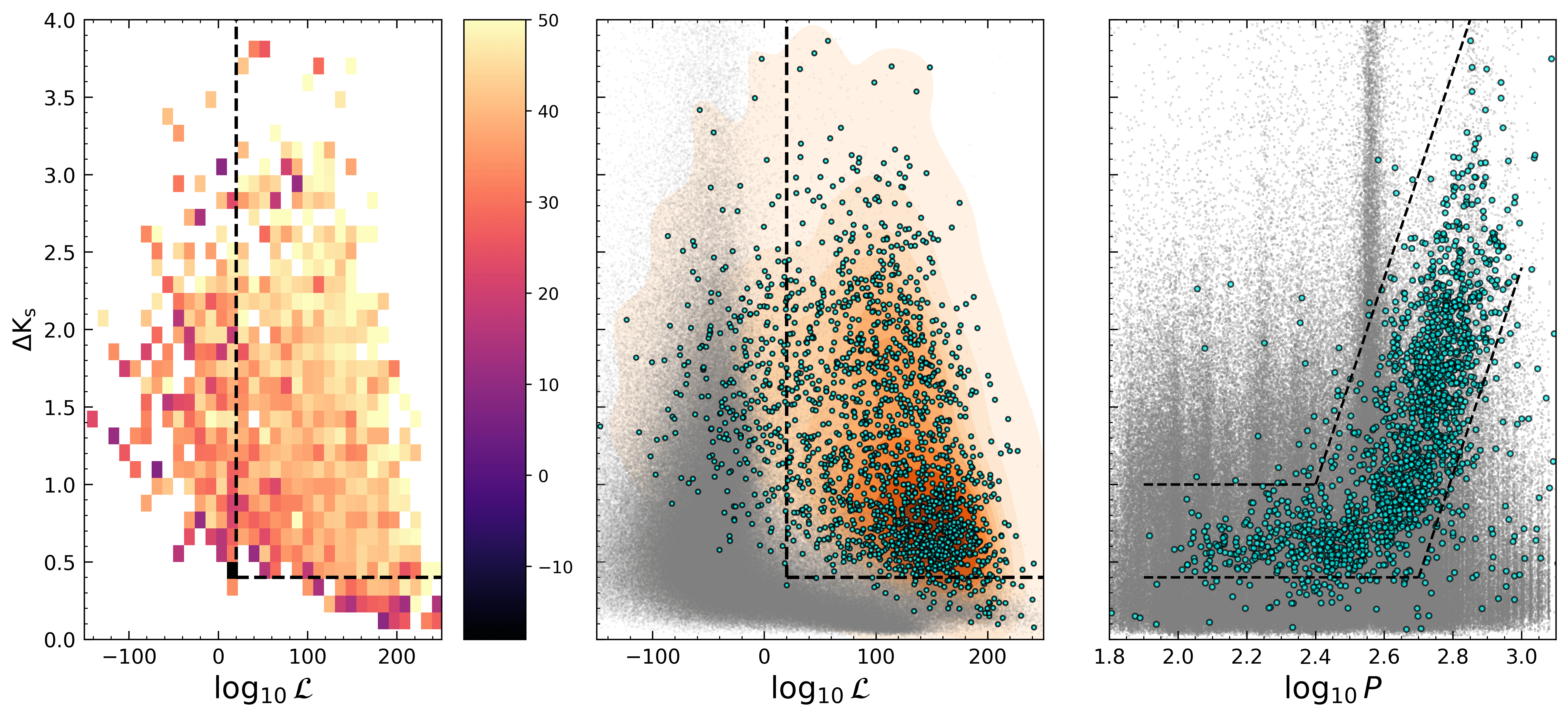}
    \caption{Distribution of VarZoo and GP candidates in likelihood, amplitude, and period space. \textit{left}: Likelihood and amplitude distribution of the VarZoo sample; the color of the bins denote the average vote score. \textit{Middle}: Distribution of Likelihood and Amplitude of the VarZoo sample and GP candidates; gray points indicate GP candidates, cyan points are the Varzoo sample, with the Orange contours indicating the density of VarZoo sample in the presented space. \textit{Right}: $\log{P}$ and amplitude distribution; black dotted lines indicate the selection borders for Mira variables.}
    \label{fig:select_func}
\end{figure*}

Mira candidates were selected based on the reported likelihood of the fit and the amplitude of the Fourier series. Fig. \ref{fig:select_func} shows the distribution of the VarZoo sample (cyan) and GP candidates (gray) in the Fourier amplitude, likelihood, and period space. For the VarZoo sample, members of the collaboration visually inspected each light curve and assigned a quality score ranging from 10 (best) to $-$10 (bad). For a given source, the total score is simply the sum of the individual scores. We bin the distribution of VarZoo stars and show the average total score per bin in the left panel of Fig. \ref{fig:select_func}. Motivated by the distribution of the VarZoo sample and the respective scores, we chose a standard cut in the amplitude of $\Delta K_{s} \geq 0.4$ and a cut in the likelihood of $\log{ \mathcal{L}} \geq 20$. We recognize that these cuts, especially the likelihood cut, remove variables that might be Miras, but have a low-quality light curve in our photometry. We choose to exclude sources with lower likelihood values as we aim to compile a high-purity catalog for variables for which we can derive reliable distances and extinctions.

Mira variables show a clear dependence between the period of the star and the amplitude of the light curve \citep[S22]{Soszynski2011, Huang2018}. Therefore, we adopt a selection box designed to broadly match the distribution of the VarZoo Miras also in the period amplitude plane ( Fig.~\ref{fig:select_func}; right panel). These cuts resulted in 1420 stars from the VarZoo sample that meet the criteria of light curve morphology and amplitude expected from a Mira variable; these stars will be referred to as VarZoo Miras. We analyze the distribution of GP candidates (gray points in Fig. \ref{fig:select_func}) and see that the vast majority of processed sources lie outside the borders of the selection box. By applying the same cuts to the output of the GP algorithm we obtain 12,346 candidates. These candidates include the 1420 stars from the VarZoo sample that meet the aforementioned criteria.

From this list of candidates, we perform a final visual inspection and also remove duplicated sources in the overlapping region between tiles. We find that 3602 sources are clearly Miras, and we can derive reliable distances and extinctions for them. These are the sources contained in the catalog released with the present paper.

\subsection{Comparison with other catalogs.}
\label{sec:Comparison_cats}

\begin{figure}
\includegraphics[width=\hsize]{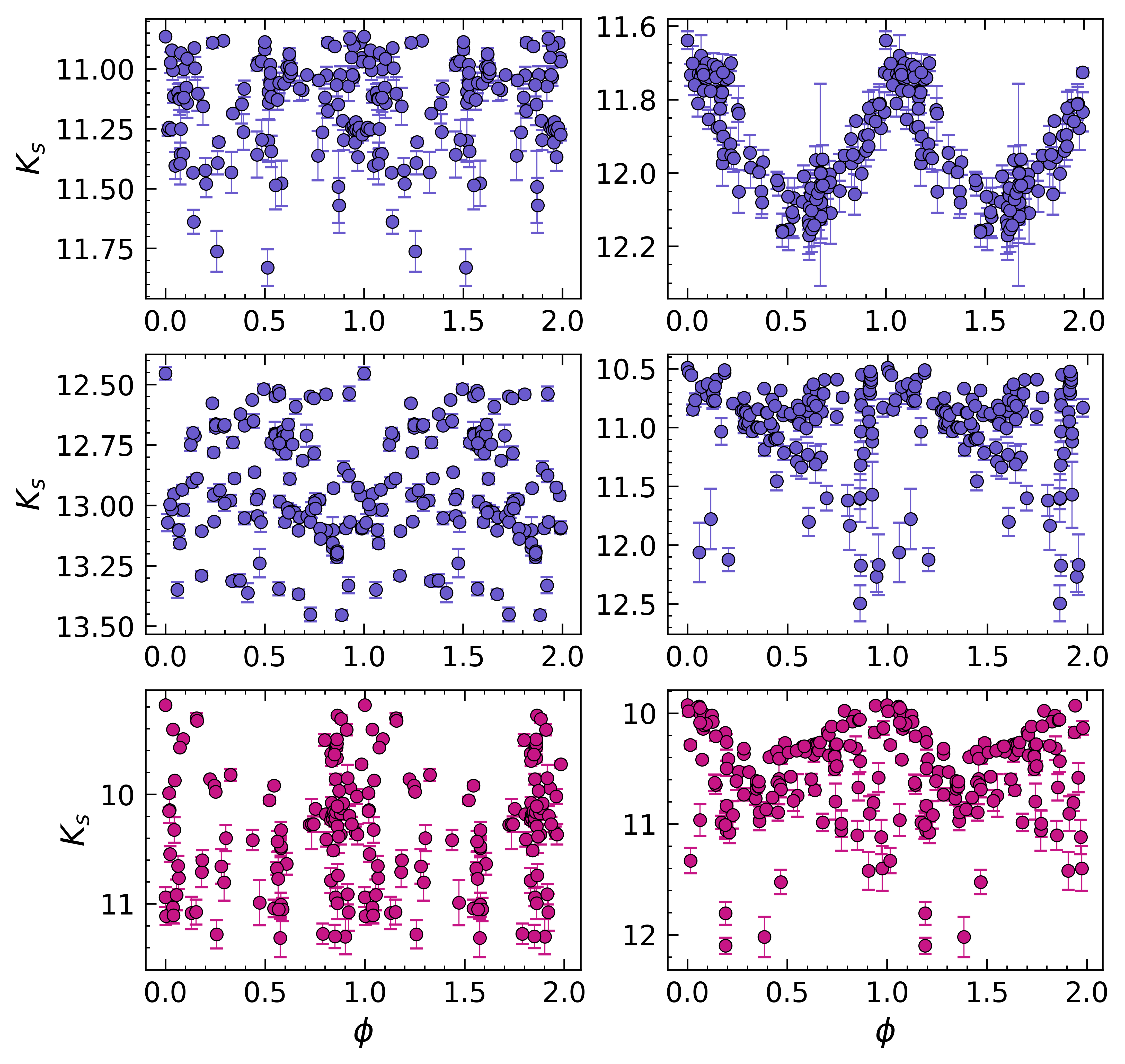}
    \caption{Example light curves for stars in other catalogs. \textit{Top and Middle panels}: light curves of stars identified as Miras in S22 but not included in the present catalog. \textit{Bottom panels}: Same as above, for sources in the catalog by \citet{Matsunaga+09}, not included in the present sample. All the light curves were folded using the period reported by the corresponding authors.}
    \label{fig:LC_comparison}
\end{figure}

The present work is based on the same images from the VVV survey as in S22, analyzed with the same PSF fitting technique. The pipeline to identify Mira variables and measure their mean magnitudes is also very similar. Therefore, a direct comparison of the results is in order. A first important difference is that the work of S22 is restricted to a 3$^\circ$$\times$3$^\circ$ region around the Galactic center, while the present work includes the whole VVV area, as shown in Fig.~\ref{fig:sky}. The catalog by S22
includes a total of 1782 Miras, while there are only 939 Miras in our catalog in the common sky region. Specifically, there are 1013 variables in S22 that were not identified as reliable Miras in the present work, whereas 170 variables that we label as Miras are not present in the S22 catalog. As for the first sample, present in S22 but not in the present work, we report a few light curve examples in Fig.~\ref{fig:LC_comparison}, all of them folded with the period reported in S22. In a few cases, like the one at the top-right corner, it is clear that we are looking at a variable star that is most likely a Mira, too. Nonetheless, this one was not fed to our algorithm because we only considered light curves whose amplitude was larger than 10 times the mean error on the data points. The vast majority of the stars in S22 not included in this work have light curves like those in the middle and top-left panels of Fig.~\ref{fig:LC_comparison}. In these cases, we might say that the variability is real, but we were unable to assign a reliable period and mean magnitude to them.

Two considerations must be kept in mind while performing this comparison. First, for the Miras in common between the two catalogs, the derived periods are virtually identical for all except $\sim$20 stars. We phase the stars with both periods, and found that only in two cases, the period
selected by our algorithm produces a folding worse than that reported by S22. These stars were removed from further consideration. Second, stars with \ks$\sim$11 mag, or brighter, are saturated in VVV. The PSF photometry performed by ALLSTAR does a good job of recovering a reasonable magnitude (see discussion in Appendix~\ref{sec:saturation_blends}, and Fig.~2 in S22). Nonetheless, we believe that a different prescription for dealing with saturated stars is most likely the reason why stars like those shown in the middle and top-left panels might have light curves compatible with Miras in the S22 photometry but not in the present one. We would also like to emphasize the importance of visual inspection of the final sample. For example, the star whose light curve is shown in the middle-right panel was included in the initial catalog, and the GP derived reasonable parameters for it. It was later discarded by visual inspection. This last step is often neglected in a world where machine learning algorithms are faster and more effective by the day. While we acknowledge that it is impossible to visually classify hundreds of thousands of light curves, we wish to stress that, although subjective, visual classification is still the most reliable method we have. When the numbers allow, this step should not be skipped.

The bottom panels of Fig.~\ref{fig:LC_comparison} show a couple of light curves for stars in the catalog by \citet{Matsunaga+09}, that were not classified as Mira in the present work. There are 85 such variables. The large majority of these stars are very bright, and the effect of saturation is stronger. It is, therefore, not surprising that we miss most of them.


\begin{figure}
\includegraphics[width=\hsize]{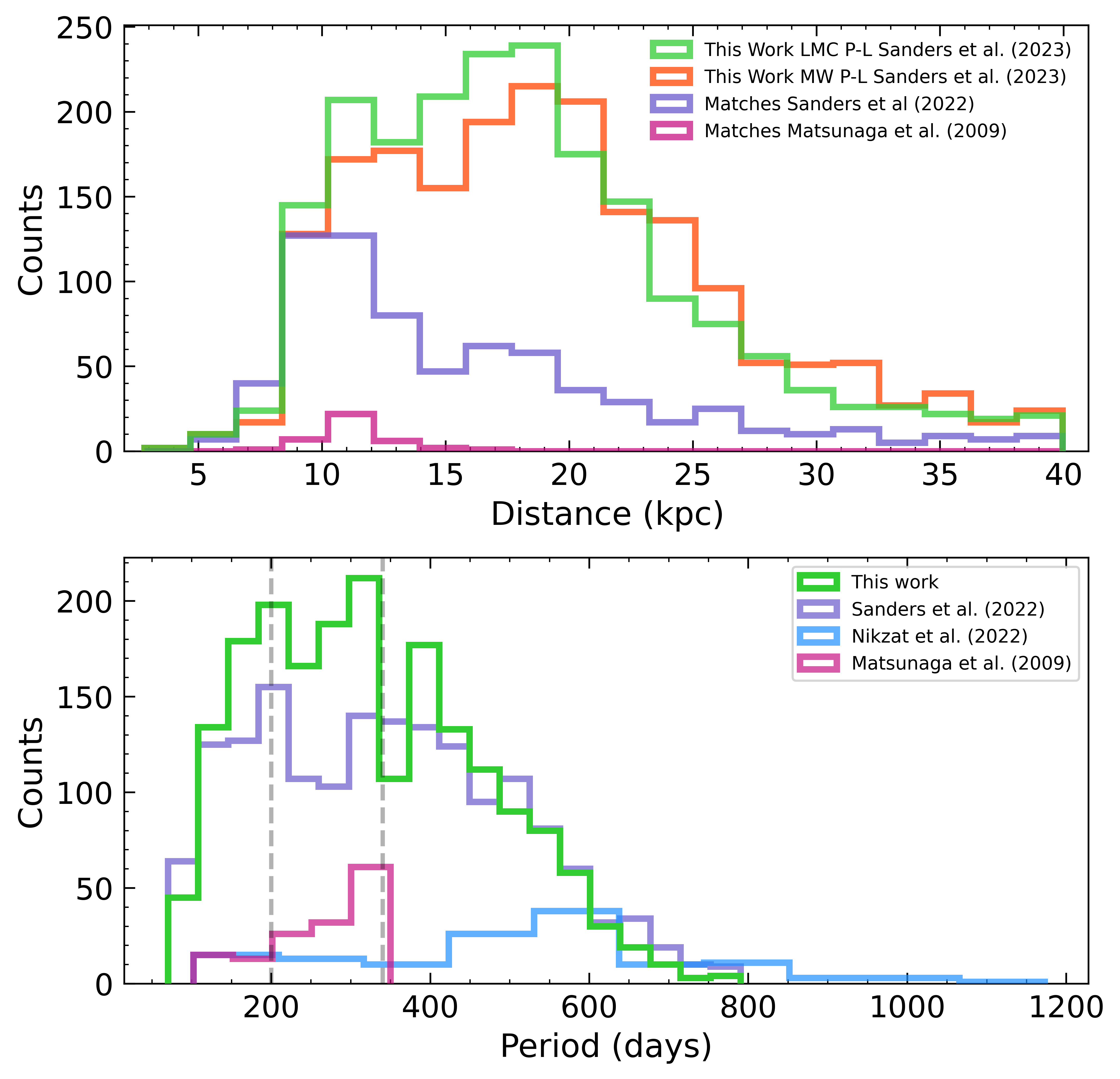}
    \caption{Comparison of different P-L relations and catalogs. \textit{Top}: Distance distribution of our sample using the P-L relations from \citet{SandersPL2023} using MW Miras (orange) and LMC Miras (green). We also show the distance distribution of matching Miras from \citet[][pink]{Matsunaga+09} and S22 (purple), also using the P-L relation using LMC Miras from \citet{SandersPL2023} adopted in this study. \textit{Bottom}: Period distribution of different catalogs. We show our final sample (green), the S22 catalog (purple), \citet{Matsunaga+09} catalog (pink), and the OGLE/VVV Mira catalog from \citet{Nikzat+2022}. We add two tentative peaks for the period distribution as gray vertical lines.}
    \label{fig:PL_comparison}
\end{figure}

Both S22 and \citet{Matsunaga+09} argue that most of the Miras in their catalogs are located very close to the Galactic center, specifically in the NSD. Unfortunately, we cannot perform a one-to-one comparison of the distances derived in the present work with those of S22, as the latter are not reported in their catalog. Still, we can compare the distance distribution derived in the two works. In the present work, we tested the P-L relation derived by \citet{SandersPL2023} from a sample of nearby Miras with Gaia DR3 parallaxes, converted to the VVV system as reported in Table~1 of S22. We also tested the P-L relation, from the same table, derived by \citet{SandersPL2023} for Miras in the LMC, assuming a distance modulus to the LMC of $\mu_{\rm LMC}$=18.477$\pm$0.026 \citep{LMC1per}. Although the two relations are very similar, the extinction law derived further below is quite sensitive to the value of the P-L coefficients, and that, of course, has a non-negligible impact on the final distance distribution. The main reason for our preference for the LMC-based P-L is the derived rotation curve, as discussed in Sect.~\ref{sec:kinematics} below. In addition, previous works have shown that the derivation of precise distances for nearby Miras is very challenging, even with Gaia DR3 parallaxes \citep[see discussion in Sect.~5.1 of][]{Huang+2024}.


The orange and green histograms in Fig.~\ref{fig:PL_comparison} show the distance distribution of the whole sample of Miras in the present work, as derived with the MW-based P-L by \citet[][orange]{SandersPL2023}, and with LMC-based P-L by the same authors, adopted here (green). The comparison shows that, had we used the MW-based P-L, we would not have found a significantly different distance distribution, especially close to the Galactic center. If anything, we would have assigned larger distances to each of our variables. Having said that, the purple histogram shows the distance distribution of the Miras in common with S22. While the distribution shows a peak at the distance of the Galactic center, we cannot confirm that most of these variables are located in the NSD due to its relatively small size and the typical distance errors. Of course, we cannot say anything about the location of the 1013 stars not included in our catalog. The very few stars that we have in common with \citet{Matsunaga+09} are also shown in Fig.~\ref{fig:PL_comparison}, and they are mostly close to the Galactic center, but they are too few to allow us to draw any conclusion.

We further contrasted our catalog with Mira catalogs based on data from the Optical Gravitational Lensing Experiment \citep[OGLE;][]{Udalski+15}. We consider the catalog from \citet{Iwanek+2022} of OGLE Miras in the Galactic bulge and disk and the VVV/OGLE catalog from \citet{Nikzat+2022}, finding 59 and 56 stars in common, respectively. As noted by \citet{Nikzat+2022}, Miras in the OGLE catalog are likely to be heavily saturated in VVV, impacting the quality of the light curve and thus hampering the ability of the GP algorithm to identify the sources.

Another important point of comparison is in the distribution of periods of the samples. We analyze the Mira period distribution of our catalog and several catalogs from the literature in the bottom panel of Fig. \ref{fig:PL_comparison}. We observe an overall uniform distribution with two small peaks at P$\sim$200 days and P$\sim$320 days for the whole sample. As recently noted by \citet{Suresh+24}, the Mira period distribution from catalogs in the literature exhibits two peaks at $\sim$200 days and $\sim$350, shown as gray dashed lines in Fig. \ref{fig:PL_comparison}. Even though our sample does not show a strong bimodality, the
two small peaks in our period distribution are broadly consistent with the expected location of the peaks from other studies, such as \citet{Matsunaga+09}, \citet{Nikzat+2022} and \citet{Suresh+24}. However, we also note that the 350-day peak is greatly affected by the adopted alias removal procedure, resulting in a gap at $P\sim 365 \ $ days, which can explain the difference between the expected and observed peaks.

\section{Sample properties}
\label{sec:SampleProp}

\subsection{Surface chemistry}

\label{sec:OC_composition}

As previously stated, AGB stars are typically categorized as C-rich or O-rich depending on the abundance ratio [C/O]. C-rich stars with [C/O]$>$1 are known as C stars, while O-rich stars with [C/O]$<$1 are named M stars, and intermediate stars with [C/O]$\sim$1 are labeled S stars \citep{Hofner+2018}. The third dredge-up process, together with the metallicity and mass of the star, determines the resulting chemical composition \citep{Gail+09}. Typically, C-rich stars are associated with metal-poor or younger systems such as the LMC \citep{Soszynski+2009}. O-rich stars, instead, are linked to metal-rich older populations like the MW bulge, although noticeably \citet{Matsunaga2017} found 5 C-rich Miras in this region. Therefore, the stellar type based on the [C/O] ratio conveys important information about the stellar content and is crucial for determining precise distances.

Many methods have been employed in the literature to distinguish between M and C stars using photometric data only. M and C type stars have been observed to exhibit different color indexes in the near-IR \citep{Feast+89, Soszynski+2009, Lebzelter+18} owing to their different compositions of circumstellar dust \citep[S22]{ItaMats11, Hofner+2016, Matsunaga2017, Hofner2022} and spectral features at several wavelengths due to C and O molecular bands \citep{SandMats2023}. For Miras in the LMC, the distinction between C-rich and O-rich chemistry has been achieved using optical and near-IR Wesenheit indexes \citep{Soszynski+2005, Soszynski+2009, Cioni+2010, Lebzelter+18}, and color index combinations \citep{Matsunaga2017, Suh+2017, Suh2018}. Furthermore, \citet{Iwanek+2021-PL} demonstrated that this distinction could also be made using only the period and a mid-IR color index. This surface chemistry distinction is not exclusive to color indexes: \citet{Cioni+03} found amplitudes of C stars to be, on average, higher than those of M stars. By using the same LMC sample of M and C type stars as \citet{Yuan+17}, \citet{Huang2018} showed that these populations can be distinct on the basis of their periods and amplitudes in the H near-IR band. Despite relatively vast literature discussing how to untangle these AGB stars, the implementation is not straightforward to apply to our sample, as large star-by-star distance spread directly affects the colors due to extinction, blending the two populations in simple color-period diagrams. 

\begin{figure}
\includegraphics[width=\hsize]{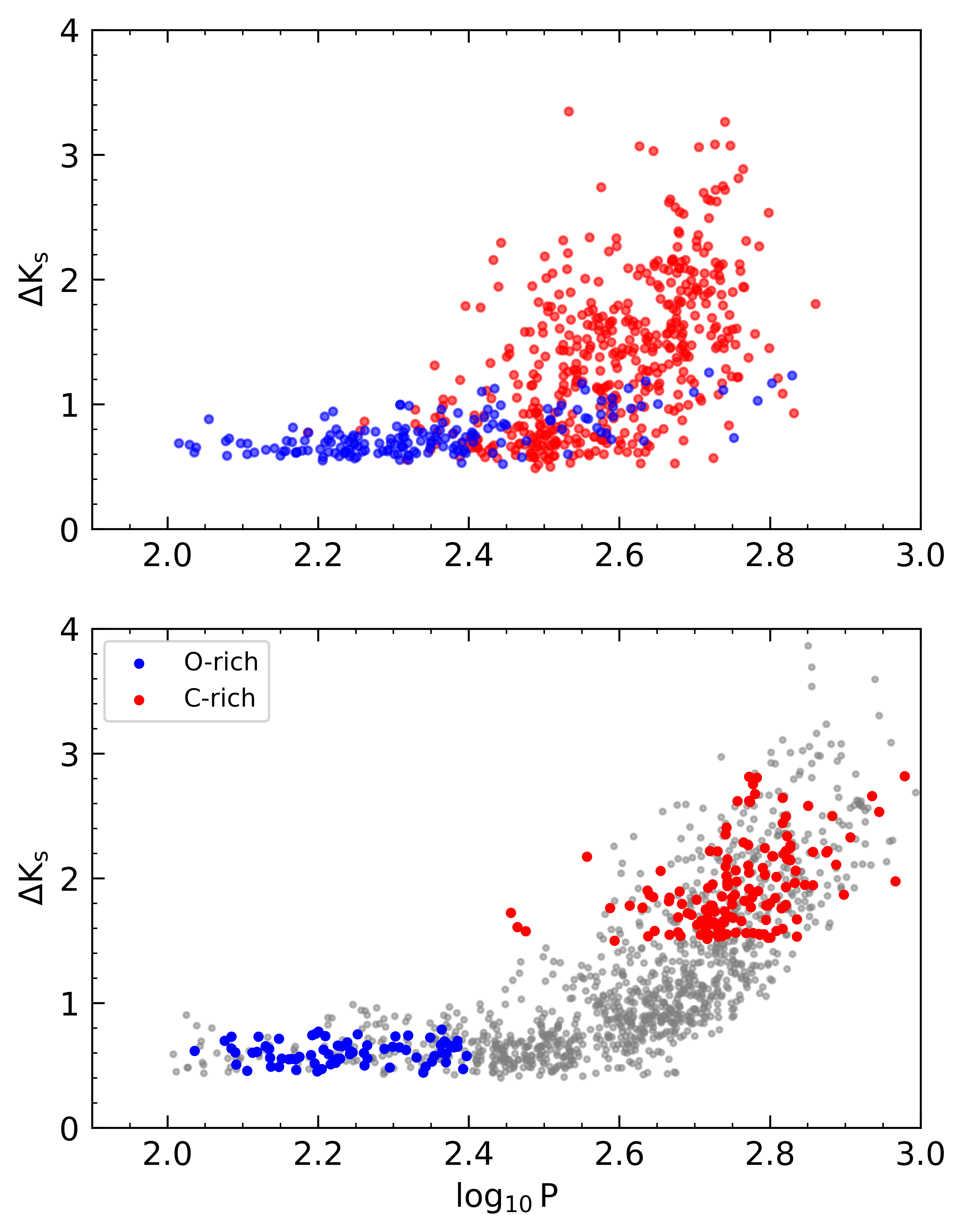}
    \caption{Period-amplitude distribution of VarZoo and LMC Miras. Red points indicate C-rich Miras, and blue points indicate O-rich Miras for both samples. \textit{Top}: LMC Miras from \citet{Yuan+17} classified using Weseinheit indexes. \textit{Bottom}: Miras from the VarZoo sample with the assigned classifications based on period and amplitude ranges.}
    \label{fig:OC_select}
\end{figure}

Distinguishing between O-rich and C-rich surface composition is crucial for adequate source-by-source reddening determination, and distances. To achieve this goal using the available photometric filters, we first explore the differences in their light curve properties. We build a preliminary sample of O-rich and C-rich Miras based on the Period vs. \kssp band amplitude measured by the Fourier fit. We restrict this analysis to VarZoo Miras only, as they have the best light curves in the sample. Motivated by the distribution of LMC Miras identified and classified as C/O rich by \citet{Soszynski+2009} and then studied in near-IR by \citet{Yuan+17}, we label as O-rich the stars with $\log_{10}(P) \leq 2.4$ and amplitudes $0.4 \leq \Delta K_{s} \leq 0.8$, and C-rich the stars with $\log_{10}(P) \geq 2.4$ and amplitudes $\Delta K_{s} \geq 1.5$ (Fig.~\ref{fig:OC_select}, bottom). Stars in the middle region can be of both types and therefore, we did not classify them into either type at this stage. By comparing to the \citet{Yuan+17} sample, shown in the top panel of Fig.~\ref{fig:OC_select}, we estimate the contamination of the O-rich sample to be $<$10$\%$, and the contamination of the C-rich sample to be negligible. 

\begin{figure}
\includegraphics[width=\hsize]{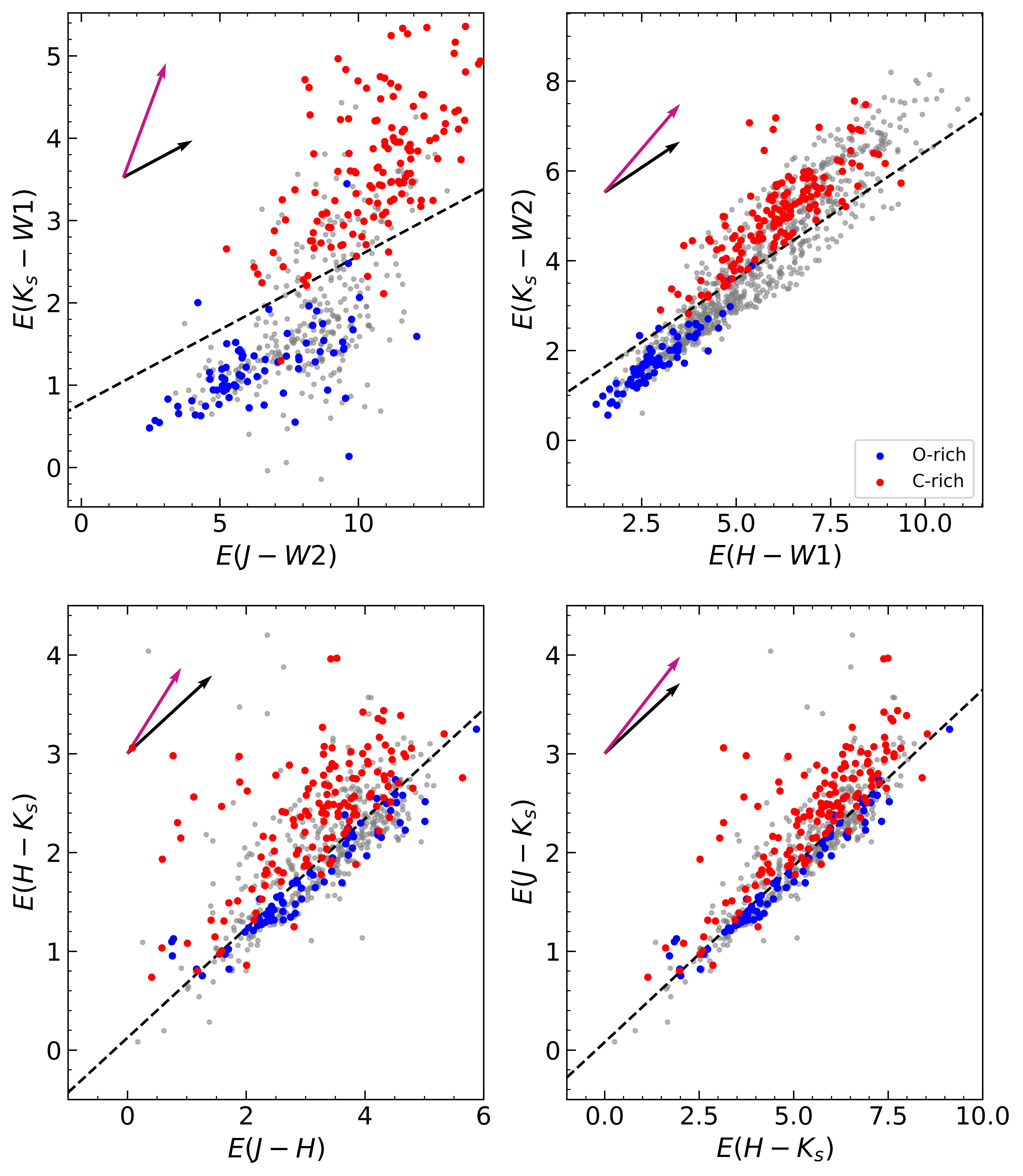}
    \caption{Distribution of C-rich and O-rich VarZoo Miras in excess color planes detailed in Table \ref{tab:ZPsOCSep}. Red points show C-rich Miras, Blue points indicate O-rich Miras, and gray points show the rest of the VarZoo sample. Black arrows indicate the extinction law from \citet{WangExtLaw19} in the top panels and this study in the bottom panels. Magenta arrows show the fit for C-rich Miras from \citet{Yuan+17} in the same color planes. Black dotted lines show the derived separations between C-rich and O-rich Miras.}
    \label{fig:OC_CC_separation}
\end{figure}

In order to see how these samples separate in the two-color planes, in Fig. ~\ref{fig:OC_CC_separation}, we plot the distribution of the preliminary subsamples of C-rich and O-rich Miras using the median magnitudes derived from the models in Sect. \ref{sec:Multiband}. We obtain the absolute magnitudes per band using the LMC O-rich P-L relation from \citet{SandersPL2023}, for the VVV filters, and the P-L relation from \citet{Iwanek+2021-PL}, for the WISE bands. Comparing the distribution of our sample with the extinction law from \citet[][black arrows]{WangExtLaw19}, we see that although our Miras show a non-negligible spread along the reddening vector, the separation between the C-rich and O-rich samples is still visible. In order to understand this separation, we perform linear fits in the same color excess planes to the LMC Mira sample from \citet{Soszynski+2009}, together with the near-IR photometry from \citet{Yuan+17} and WISE/Spitzer data by \citet{Iwanek2021}. As noted by \citet{Yuan+2018}, C-rich Miras from the LMC occupy a distinct sequence with a slope (magenta arrows) different from the Galactic reddening vector, hence allowing us to disentangle the reddening for these sources. To separate the samples, we use a linear cut with the slope of the \citet{WangExtLaw19} extinction law and a zero point that maximizes the \textbf{F1} score, defined as:

\begin{equation}
    \textbf{F1} = \frac{2 \cdot \textbf{TP}}{2 \cdot \textbf{TP} + \textbf{FP} \cdot \textbf{FN}}
\end{equation}
\noindent
\textbf{TP} are the True Positives (O-rich Miras classified as such), \textbf{FP} are the False Positives (C-rich Miras labeled as O-rich), and \textbf{FN} are the False Negatives (O-rich Miras labeled as C-rich). The \textbf{F1} score adopts values from 0 to 1, indicating a complete misclassification and a perfect classification, respectively. This cut, using WISE filters and the extinction law from \citet{WangExtLaw19}, allowed us to isolate stars with clear excess in color that separate from the O-rich region without a previous assumption of an extinction law in the VVV bands. As shown in Fig.~\ref{fig:OC_CC_separation}, the separation is maximal in the top left panel. Nonetheless, we applied the same process also to a combination of only VVV filters, because not all the stars in the sample were identified in the WISE photometry. We repeated the procedure using the extinction law derived in Sect. \ref{ExtLaw}; the results for the zero points obtained for the different color planes are detailed in Table \ref{tab:ZPsOCSep}. O-rich stars tend to follow narrower sequences in the color planes, aligned with the interstellar extinction vector due to their optically thin envelopes, whereas C-rich stars extend to redder colors with a larger scatter. This effect is more notorious for WISE filters, as flux contributions from the circumstellar shell become more important in the mid-IR for C-rich Miras, whose spectral energy distribution can be described by a double Planck law \citep{Iwanek2021}. To classify Mira candidates by surface chemistry in our catalog, we used the above color planes, favoring the use of WISE data where available. We prioritized color planes by the angular difference between the circumstellar extinction vector (magenta arrows in Fig. \ref{fig:OC_CC_separation}) and the interstellar extinction vector (black arrow in Fig. \ref{fig:OC_CC_separation}). This order is also showcased in Table \ref{tab:ZPsOCSep}. It is important to note that our classification method requires measurements in at least three or four bands, depending on the color plane used. However, due to the high extinction affecting the J  and H  bands and the blending issues with WISE photometry near the Galactic plane, 512 sources in our catalog lack the necessary filter combinations. As a result, we could not provide a classification for these sources.

\begin{figure}
\includegraphics[width=\hsize]{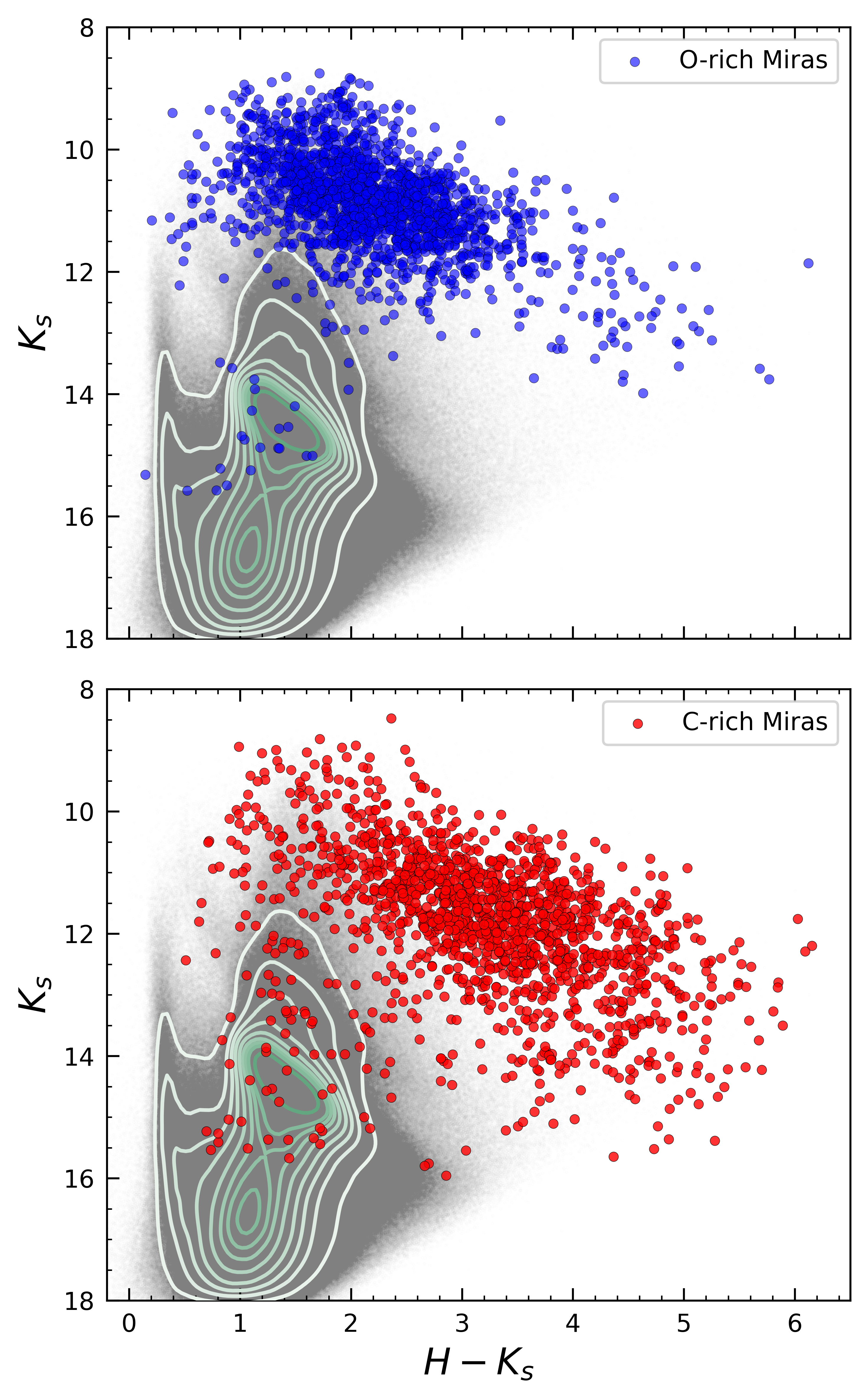}
    \caption{CMD using H and \kssp of Miras in our sample. \textit{Top}: CMD of O-rich Miras. \textit{Bottom}: CMD of C-rich Miras. Both panels include a sample of stars from VVV as gray points, with the corresponding density contours in green. In this CMD, the extension of C-rich Miras to redder colors is clear.}
    \label{fig:CMD}
\end{figure}

We show the distribution of C-rich and O-rich for stars of our Mira sample in a color-magnitude diagram (CMD) in Fig. \ref{fig:CMD}. We clearly see that C-rich Miras have, on average, redder colors, owing to their optically thick dust envelopes. 

\begin{table}
 \caption{Separating lines for O-rich and C-rich Miras in the Color Excess planes considered.}
 \label{tab:ZPsOCSep}
 \begin{tabular}{lllll}
  \hline
  CE X & CE Y & Slope & Zero-Point & F1\\
  \hline
    E(J - W2) & E(\ksnosp - W1) & 0.179 & 0.770 & 0.953\\[2pt]
    E(H - W1) & E(\ksnosp - W2) & 0.565 & 0.770 & 0.952\\[2pt]
    E(J - H)  & E(H - \ksnosp)  & 0.554 &  0.122 & 0.830\\[2pt]
    E(H - \ksnosp) & E(J - \ksnosp) & 0.356 & 0.078 & 0.830\\[2pt]
    \hline

 \end{tabular}
\end{table}

\subsection{Extinction law}
\label{ExtLaw}

In order to derive distances and reddening for each source, again, we obtain the absolute magnitudes in the VVV filters using the near-IR P-L relation by \citet{SandersPL2023} using LMC Miras in the VVV photometric system, detailed in Sect. \ref{sec:3Ddistrib}. For distance purposes, we focus on the \kssp band as it has the highest sampling and a lower typical error. To obtain the total extinction in the \kssp band ($A_{K_{s}}$) from the observed color excess, we need an accurate value of the total-to-selective extinction ratios in a filter combination. As discussed in \citet{MatsunagaCCs+2016} in the case of Classical Cepheids, and in \citet{Nikzat+2022} in the case of Miras, differences in the adopted extinction law directly impact the determined distances, especially in the highly obscured regions close to the Galactic plane. The values of the selective-to-total extinction ratios $A_{K_{s}}/E(H - K_{s})$ in the literature span a wide range of values, from 1.10 \citep{AlonsoGarcia2017} to 1.61 \citep{Nishiyama2009}. In the case of $A_{K_{s}}/E(J - K_{s})$, the range is smaller, from 0.428 to 0.528, drawn from the same two references. Using Classical Cepheids at the disk far side, \citet{Minniti+2020} determined a value of 1.308 and 0.465, for $A_{K_{s}}/E(H - K_{s})$ and $A_{K_{s}}/E(J - K_{s})$ respectively. Due to this non-negligible spread, it is important to determine an extinction law suitable for our sources.

To determine the extinction law for our sample of Miras, we used O-rich Miras from the VarZoo sample, cleaned from C-rich contaminants, using the WISE color excess planes and the \citet{WangExtLaw19} extinction law (Fig.~\ref{fig:OC_select} top), as detailed in Sect. \ref{sec:OC_composition}. We also imposed additional cuts in likelihood $ \log_{10}{\mathcal{L}} \geq 50$ to only consider the highest quality stars. As mentioned, Miras can show trends in their average magnitudes over several cycles \citep{Uttenthaler+2011}. From the modeling described in Sect. \ref{sec:Identification}, we defined the parameter \textbf{$\rm MinMax K_s$} to describe the absolute difference in magnitude of the polynomial fitting the median trend of the light curve. For this sub-sample, we impose MinMaxK$_{\rm s}$$\leq$0.05 mag to include only light curves that have a stable mean brightness over the time baseline of the survey. We performed a linear fit to the distribution of O-rich Miras in the color excess planes, imposing a fixed intercept of 0 and a $3\sigma$ clipping to exclude outliers.
From this, we obtained the following mean color excess ratios:

\begin{equation} \label{eqn:JKHK}
    \frac{E(J -K_{s})}{E(H - K_{s})} = 2.801 \pm 0.016,
\end{equation}

\begin{equation} \label{eqn:JHJK}
    \frac{E(J -H)}{E(J - K_{s})}     = 0.6438 \pm 0.0021.
\end{equation}

\noindent
Assuming that the extinction law in the near-IR bands follows a power law  $A_{\lambda}$$\propto$$\lambda^{-\alpha}$, we can get the power law index from the mean color excess ratios. For the VISTA bands, the effective wavelengths ($\lambda_{\rm{eff}}$) of each filter are 1.254 $\mu m$ (J), 1.646 $\mu m$ (H), and 2.149 $\mu m$ (\ksnosp). From (\ref{eqn:JKHK}) and (\ref{eqn:JHJK}) we obtained $\alpha = 2.11 \pm 0.02$. The derived selective-to-total extinction ratios are:

\begin{equation} \label{eqn:AKJK}
    \frac{A_{K_{s}}}{E(J - K_{s})} = 0.471 \pm 0.010,
\end{equation}

\begin{equation} \label{eqn:AKHK}
    \frac{A_{K_{s}}}{E(H - K_{s})} = 1.320 \pm 0.020.
\end{equation}

\noindent
From these values and the measured color excess, we can calculate $A_{K_{s}}$ for our Mira sample.

\section{Three-dimensional spatial distribution}
\label{sec:3Ddistrib}

\subsection{Distance determination}
\label{sec:distances}
\begin{figure*}[htb!]
\includegraphics[width=1.035 \hsize]{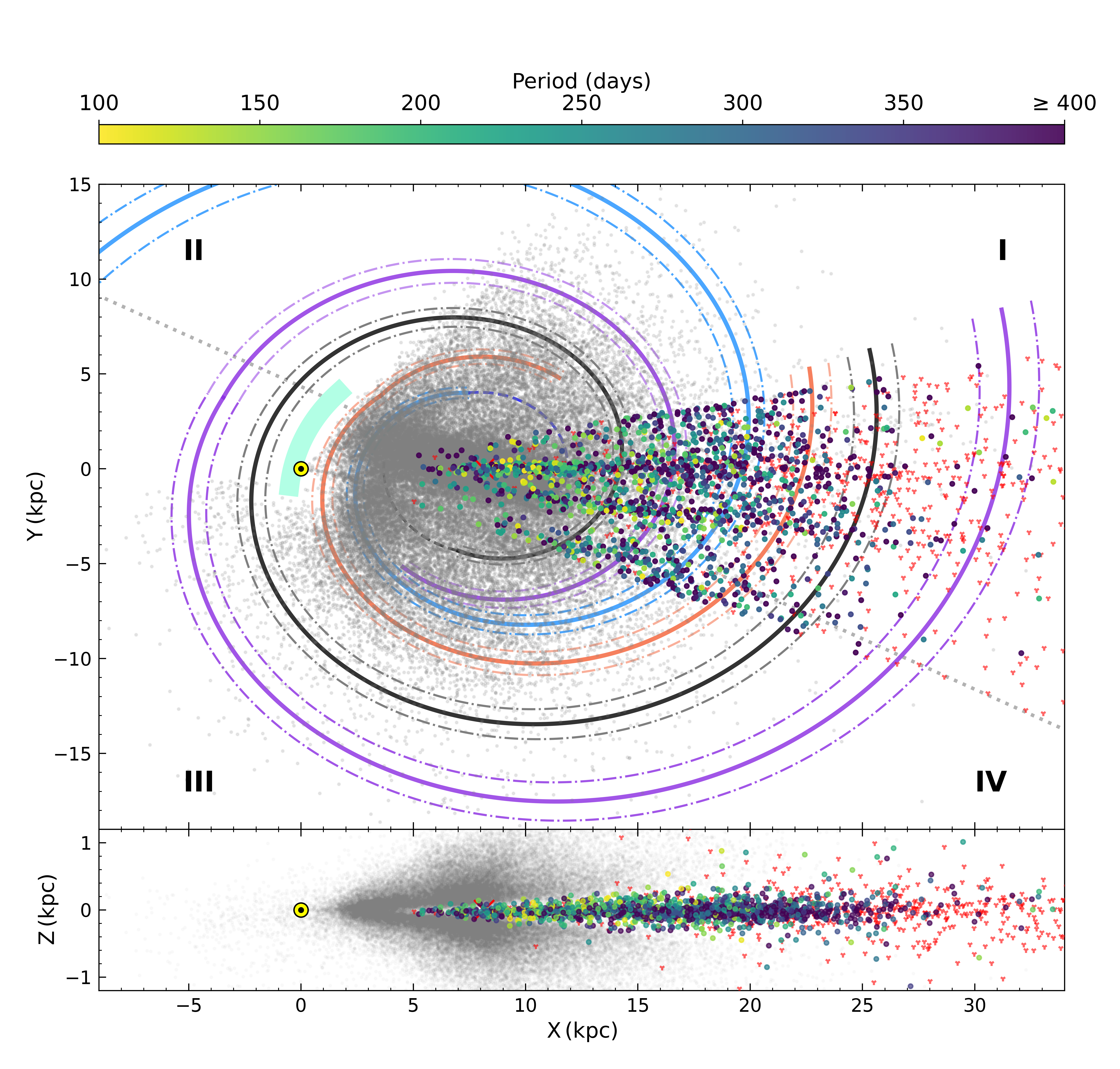}
    \caption{Three-dimensional distributions of Miras in the MW. The position of O-rich stars in our sample is shown in circles color-coded by period. C-rich stars are included as red ticks. We over-plot the Miras from OGLE project presented in \citet{Iwanek+2022} with the distances reported in \citet{Iwanek+23} as gray dots. The Sun is located at (X,Y)=(0,0), and marked with a yellow dot, and the Galactic center is marked with a red cross at (X,Y)=(8.34,0). Log-periodic spiral arm segment fit of the local arm based on VLBI trigonometric parallaxes of high-mass star-forming regions by \citet{Reid+19_SF_parallax} is represented with a turquoise-colored section. The thick lines represent qualitative logarithmic spiral arm fits from \citet{Minniti+2021}, with the thin dashed lines marking the $1\sigma$ arm widths reported in \citet{Reid+19_SF_parallax} (Sct-Cen arm, blue; Sgr-Car arm, orange; Perseus arm, black; Outer arm, purple). The line of nodes of the Galactic warp model by \citet{skowron19} is also shown for reference (gray dotted line). The bottom panel shows the XZ view of the Mira distribution in the MW.}
    \label{fig:distances}
\end{figure*} 

To obtain distances to individual stars, we used the average \kssp magnitude derived in Sect. \ref{sec:Identification} and Sect. \ref{sec:Multiband}, together with the extinction law derived in Sect. \ref{ExtLaw}. The distances presented here are based on the H and \kssp band magnitudes and selective-to-total extinction ratios. We chose this combination of filters because these bands have lower typical errors in average magnitude, and the H and \kssp based extinction ratios since $\sim$38$\%$ of the sample do not have J band measurements. We calculate the distance modulus using the expression

\begin{equation}
\label{eqn:mod}
    d =  10^{1 + 0.2 (K_{s} - A_{K_{s}} - M_{K_{s}}) },
\end{equation}

\noindent
where \kssp is the intensity-based mean magnitude in the \kssp band derived from the modeling described in Sect. \ref{sec:Identification}, $A_{K_{s}}$ is the extinction in the \kssp band, and $M_{K_{s}}$ is the absolute magnitude obtained from the P-L relation. As mentioned in Sect. \ref{sec:Comparison_cats}, we use the P-L relation from \citet{SandersPL2023} using LMC Miras. The adopted P-L relations were transformed to the VVV photometric system using the coefficients from \citet{gf+18_VVVcal}, obtaining

\begin{equation}
    \resizebox{0.93 \columnwidth}{!}{$
    \begin{aligned}
    M_{J}     &= -6.30 - 3.32 \ (\log_{10} {P}-2.3) - 7.29 \ \mathcal{H}_{2.6} \  (\log_{10}{P} - 2.3)^{2},  \\
    M_{H}     &= -6.67 - 3.33 \ (\log_{10}{P}-2.3) - 6.85 \ \mathcal{H}_{2.6} \ (\log_{10}{P} - 2.3)^{2},  \\
    M_{K_{s}} &= -7.01 - 3.73 \ (\log_{10}{P}-2.3) - 6.99 \ \mathcal{H}_{2.6} \ (\log_{10}{P} - 2.3)^{2},  \\
    \end{aligned}
    $}
    \label{eqn:P-Ls_VVV}
\end{equation}
    
\noindent
here, $\mathcal{H}_{2.6}$ indicates a simple Heaviside function, that takes value 1 for $\log{P} \geq 2.6$ and 0 otherwise. The statistical uncertainty of the distance can be derived by applying the propagation of the errors to Eq. (\ref{eqn:mod}):

\begin{equation}
\label{eqn:mod_err}
    (\Delta d)^{2} = (0.46d)^{2}(\delta K_{s}^{2} + \delta A_{K_{s}}^{2} + \delta M_{K_{s}}^{2}),
\end{equation}

\noindent
here $\delta K_{s}$ is the mean magnitude error in the \ksnosp-band, $\delta A_{K_{s}}$ is the extinction error, and $\delta M_{K_{s}}$ is the absolute magnitude error. The value of  $\delta K_{s}$ was obtained from the sampling of the median model derived in Sect. \ref{sec:Multiband}, and $\delta M_{K_{s}}$ was derived from the error propagation of the P-L relation and the reported coefficients from \citet{SandersPL2023}. We note that the term $\delta A_{K_{s}}$ is the largest source of errors due to the large color excess in the sample and the larger uncertainties of the extinction law. We note that, as P-L relations, extinction law, and modeled median magnitudes in the WISE bands have been used in this study, it is also possible to use WISE filters for distance determination. In Appendix \ref{sec:WISE_distances}, we determined distances using only WISE filters by following the same procedure detailed in the current section and compared these distances with our adopted values. We found that the WISE-based distances are in good agreement with the adopted distances, but exhibit significantly larger errors. We attribute this increased uncertainty to the larger PSF of WISE, causing blending effects near the Galactic plane, as well as uncertainties in the extinction law. Therefore, we favor the use of \ksnosp-band-based distances in the following sections.

In Appendix \ref{sec:catalog}, we present the header of the final catalog, available at the CDS.

\begin{figure}
\includegraphics[width=\hsize]{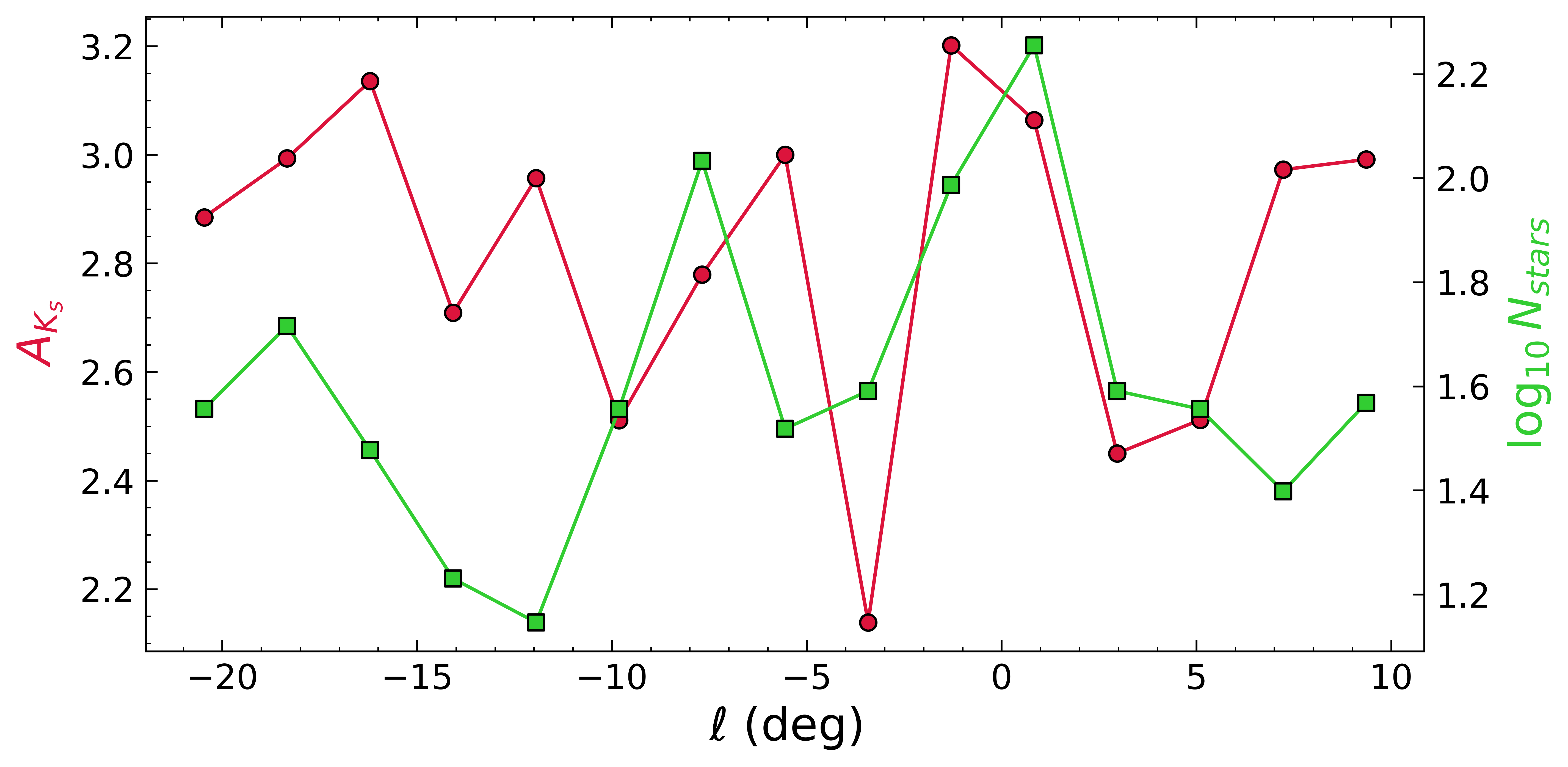}
    \caption{Correlation between stellar counts and extinction. We make equally spaced bins using the Galactic longitude and show the average extinction in the \kssp band (red) and the logarithm of the number of Miras per bin (light green).}
    \label{fig:ext_glon}
\end{figure}

\subsection{3D distribution}


The distribution of our Miras, projected to the plane of the Galaxy, is shown in Fig. \ref{fig:distances}. O-rich Miras are shown as dots, color-coded by their respective period. We note that Miras with periods longer than $\sim$400 days are often neglected from distance analysis due to the onset of Hot Bottom Burning (HBB) at the corresponding mass range, as well as the increasing mass loss rates expected for more massive stars \citep{Uttenthaler+15, Uttenthaler+19}. These processes increase the scatter from the P-L relation at higher periods, thus making distances less reliable. Here, we report distances to these sources, but they are not included in the analysis. We choose to also include C-rich Miras in the distribution (red ticks in Fig. \ref{fig:distances}), although their distances are likely over-estimated due to additional reddening from the circumstellar shell. 


A clear feature in Fig. \ref{fig:distances} is the presence of sequences along lines of sight with higher stellar density. To explain the origin of these features, we plot in Fig. \ref{fig:ext_glon} the average extinction in the \kssp band (red) and the number of sources (green) in longitude bins. We see a clear correlation between the two curves, showing that the number of detected Miras is larger for lines of sight-piercing regions with higher average extinction. We interpret this as due to low-extinction Miras being saturated in VVV.

Even though saturation effects are expected for bright sources at \ksnosp $\leq 12$ \citep{contrerasramos17}, \citet{Sanders2022} showed that no significant bias in the saturation regime was observed for Miras in the VIRAC2 catalog \citep{Virac}. Given that both the VIRAC2 catalog and the catalog from \citet{contrerasramos17} are based on the same VVV images using PSF-based methods, we consider that no significant biases in the magnitudes are present for our selected sources, and thus, the provided distances are reliable. We expand on this point in the Appendix~\ref{sec:saturation_blends}.

\begin{figure}
\includegraphics[width=\hsize]{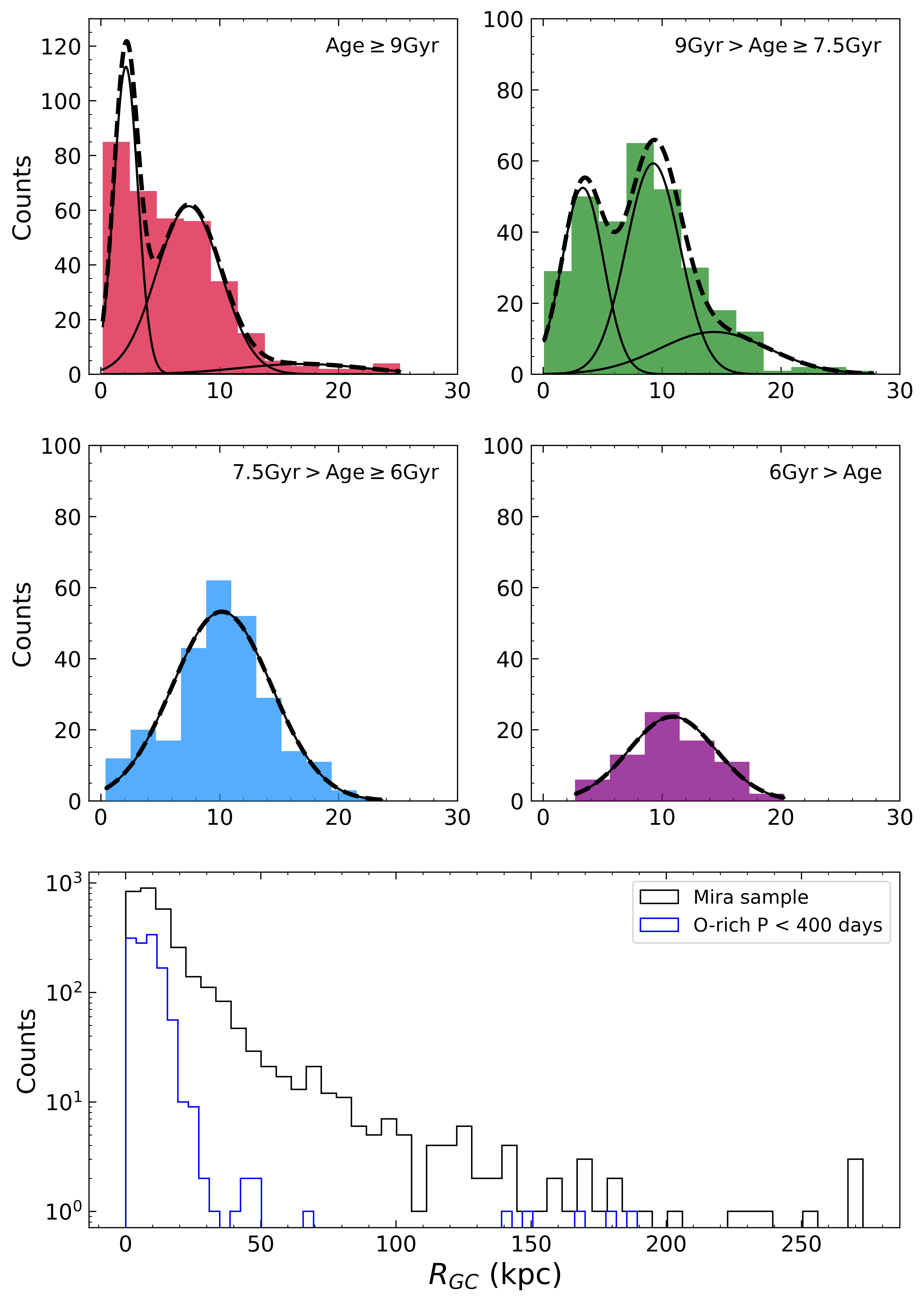}

    \caption{Histograms of galactocentric distance (R$_{\rm GC}$) binned by age.
    \textit{Top and middle panels}: Radius of stars in the Galaxy binned by age. The age range decreases from top to bottom and left to right. Black dashed lines represent the Gaussian mixture model fitted to the distribution. Individual components are presented as bold black lines. \textit{Bottom}: Full histogram of galactocentric distances for the full sample (black) and the O-rich Miras with periods under 400 days (blue).}
    \label{fig:insideout}
\end{figure}

In order to explore the correlations between the position of the Miras in the Galaxy and their age, we use the Period-Age relation of \citet{Zhang+2023} and a Sun-Galactic center distance of R=8.34 kpc \citep{Reid+14_GC_dist}. To get precise distances, we limit the analysis to O-rich stars with errors in distance less than 10\% and periods less than 400 days, resulting in 972 total sources. We bin the data in four age bins, as shown in Fig. \ref{fig:insideout}. To understand how the distance distribution changes with age, we run a grid of Gaussian mixture modeling with a maximum of 3 components, choosing the number of components that minimizes the AIC. For the two oldest bins, we observe three components, where the two furthest components increase both in weight and distance as age decreases. For the two youngest bins, the distribution is best represented by a single Gaussian. We find that the position of the component with the highest weight is located at R$_{\rm GC} = 7.40\ \rm{kpc}, 9.25 \ \rm{kpc}, 10.19 \ \rm{kpc}, 10.88 \ \rm{kpc}$ as age decreases, pointing to a reverse correlation between the age of the stars and their radius in the Galaxy. Although saturation effects decrease the observed number of Miras at low R$_{\rm GC}$, this effect should mainly impact the component closest to the Galactic center and have a slight effect at higher R$_{\rm GC}$. This is indeed seen in the decreasing weight of the innermost component as age decreases, with it completely disappearing for the youngest bins. Note that the absolute count per bin is not directly related to the underlying distribution of stars as a function of R$_{\rm GC}$, as the innermost bins include stars closer than the GC and east/westwards of it, while at larger R$_{\rm GC}$ we are only sampling a narrow cone. Despite the observational effects influencing the distribution of stars, the scenario described by the correlation between the position and age of stars in the sample is consistent with the inside-out formation of the Galactic disk, where star formation started in the inner regions of the MW and gradually moved outwards \citep{Chiappini+97, Schonrich+17_insideout}.
The presence of a centrally confined component traced only by old Miras ($\gtrsim$8 Gyr) is also consistent with a dominantly old component in the Galactic bulge \citep[see][and references therein]{Joyce+23}.

In the bottom panel of Fig. \ref{fig:insideout}, we see the distribution of distances for the full sample (black) and the O-rich Miras with periods shorter than 400 days (blue). We see that the vast majority of the sample is located well within the Galactic disk, with the noticeable exception of $\sim 10$ distance indicator Miras. Miras at these distances were also found by \citet{Nikzat+2022} using VVV/OGLE data and are likely members of substructure within the Galactic halo, such as unknown globular clusters, dwarf spheroidal galaxies, or even the Sagittarius dwarf spheroidal galaxy \citep{Ibata94}.

\begin{figure}
\includegraphics[width=\hsize]{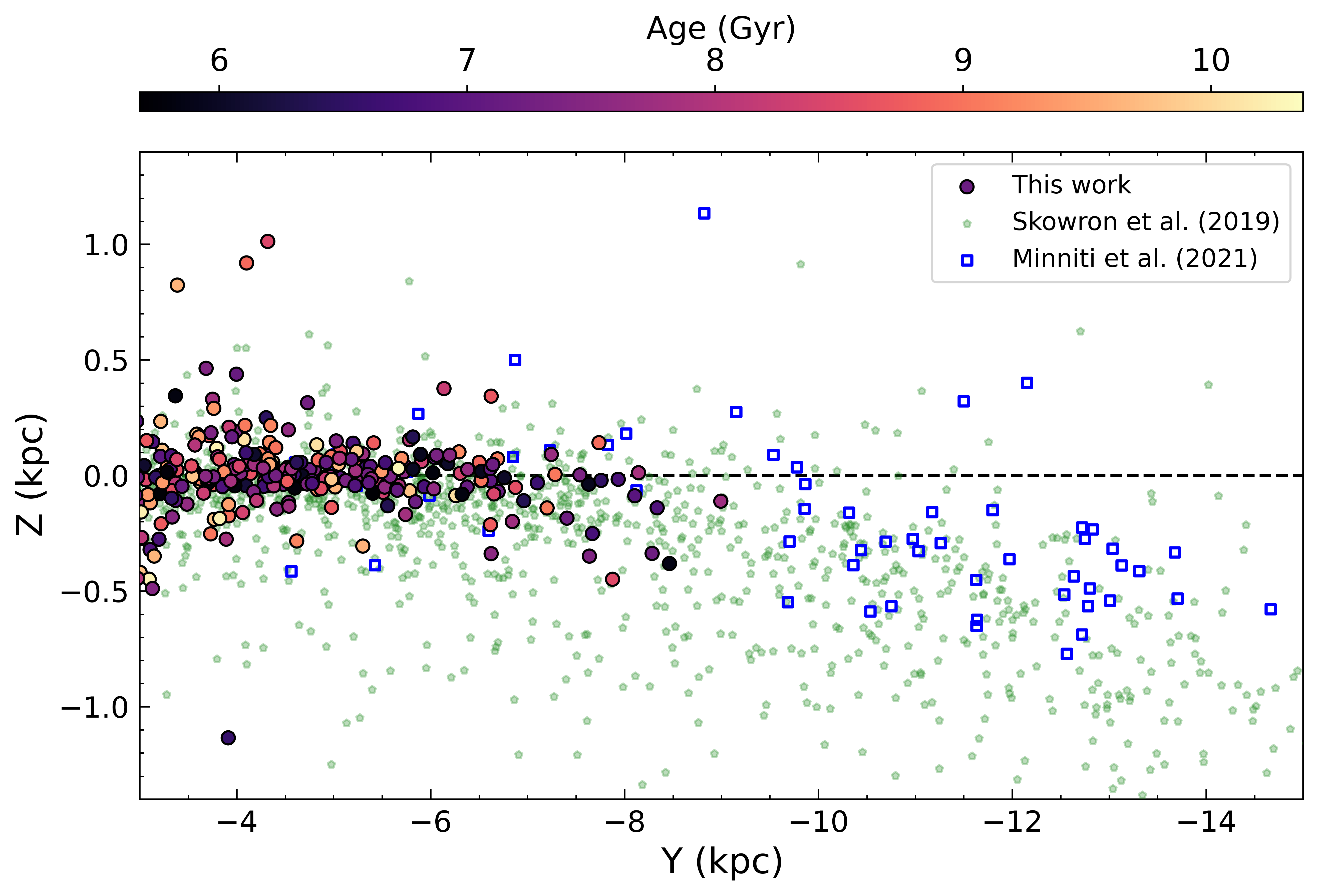}

    \caption{Projection of our Mira sample in Galactic Y versus Z coordinates. Circles represent Miras color-coded by their estimated age; We include the Classical Cepheids samples from \citet[][green polygons]{skowron19} and \citet[][blue squares]{Minniti+2021}. We include only sources in the respective samples inside the region $-15 \ \rm{kpc} \lesssim Y \lesssim -5 \ \rm{kpc}$ and $-1.5 \ \rm{kpc} \lesssim Z \lesssim 1.5 \ \rm{kpc}$.}
    \label{fig:warp}
\end{figure}

As shown in Fig. \ref{fig:distances}, a significant number of Miras lies in the region affected by the warp of the Galactic disk, from $-15 \ \rm{kpc} \lesssim Y \lesssim -5 \ \rm{kpc}$ and $-1.5 \ \rm{kpc} \lesssim Z \lesssim 0 \ \rm{kpc}$. In Fig. \ref{fig:warp} we explore the position of Miras in our catalog in that region. We find that Miras do follow the warping of the Galactic disk, aligned with the distribution of Classical Cepheids located in the region. We do not find any correlation between the Mira position in the Galaxy and star-forming regions such as the spiral arms, likely due to their ages ($\gtrsim 5 \rm Gyr$) and migration effects inside the Galaxy.

\section{Kinematics}
\label{sec:kinematics}

\begin{figure}
\includegraphics[width=\hsize]{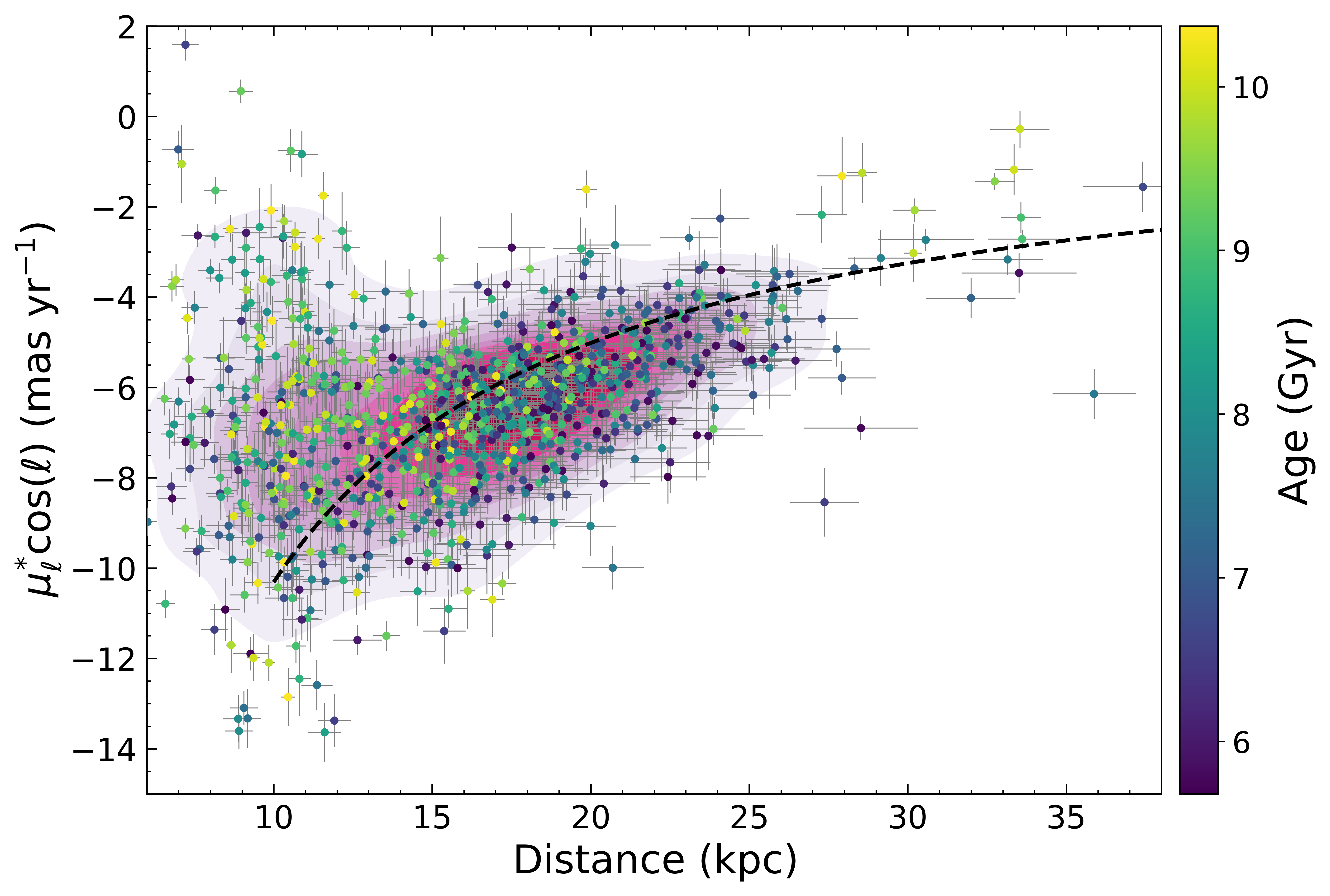}

    \caption{Rotation curve in proper motion space. O-rich Miras in our sample are shown as circles color-coded by age. Pink contours show the density percentiles. The dashed line represents a model of the rotation curve from  \citet{mroz+19} as seen for a line of sight through the Galactic center.}
    \label{fig:rotation_curve_pms}
\end{figure}

In this section, we analyze the kinematic footprint of our Mira sample using the VVV PM data. The distribution of longitudinal proper motions ($\mu^{*}_{\ell}$ $\cos$b), calibrated to the Gaia reference system and projected to a line of sight through the Galactic center, is shown in Fig. \ref{fig:rotation_curve_pms}, as a function of the heliocentric distance. Stars in the Bulge region have a wider spread in proper motions consistent with older bulge populations \citep[see Fig. 5 from][]{JOlivares+24}. As the distance increases, a transition to disk-like kinematics is observed, following the model by \citet{mroz+19} based on the near disk and projected to a line of sight through the Galactic center.

\begin{figure}
\includegraphics[width=\hsize]{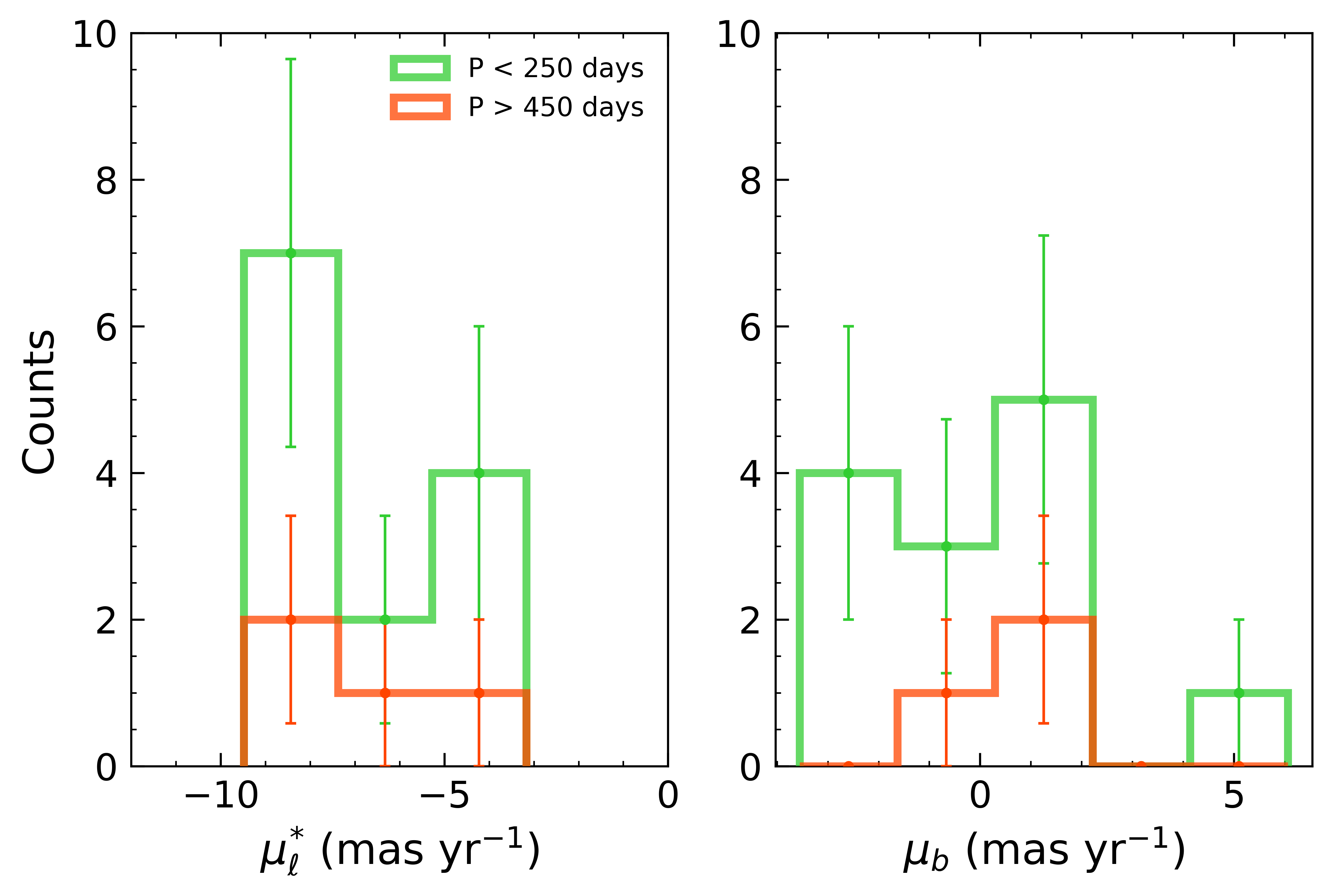}
    \caption{Histogram of PMs for stars consistent with the NSD position. \textit{Left}: Histogram of longitudinal proper motions. Green lines correspond to Miras with periods in the range 100~days$\leq$P$\leq$250 days and orange to P$\geq$450 days. The Galactic center has $\mu_{\ell}$=$-$6.41 mas $\rm{yr}^{-1}$. \textit{Right}: Same as left panel for $\mu_b$.} 
    \label{fig:NSD_hists}
\end{figure}

As mentioned in Sect. \ref{sec:Comparison_cats}, we now investigate the signal of the NSD as seen by our sample of Miras and determined distances. In Fig. \ref{fig:NSD_hists} we present the histograms of PMs of stars in our sample with distances consistent with the projected position of the NSD in the Galaxy. Following Fig. 7 from \citet{Sanders+2024}, we apply cuts in Galactic coordinates $|\ell| \leq 0.4^{\circ}$ and $|b| \leq 0.4^{\circ}$ and study two period ranges from $100 \, \rm{days} \leq \rm{P} \leq 250 \, \rm{days}$ and $450\, \rm{days} \geq \rm{P}$. Additionally, we impose a cut in R$_{\rm GC} \leq 1 \ \rm{kpc}$ to isolate the signal of the NSD. It is important to notice that this cut far exceeds the usual radius of the NSD from \citet{Sormani+22} of $R_{\rm{NSD}} \lesssim 300 \ \rm{pc}$, yet we found it necessary to include a minimum amount of stars for a reasonable comparison, and this value is still consistent with the model from \citet{Debattista18}. From Fig. \ref{fig:NSD_hists}, we observe a slight double-peaked distribution for the older period range and no peaks for the youngest Miras in the period ranges proposed in \citet{Sanders+2024}. This slight bimodality for Miras of periods shorter than 250 days cannot be explained by a rotating structure such as the NSD, as the corresponding distribution in $\mu_{b}$ is not consistent with disk-like kinematics. We conclude that in the Mira sample compiled and analyzed in this work, no strong signal of the NSD is observed.

%

To further inspect the kinematics of Miras in the bulge and far disk, we calculate the transverse velocities as detailed in \citet{Du+2020}:

\begin{equation}
    v_{\ell, HC}^{*} = 4.74 \frac{\mu_{\ell}^{*}}{\rm{mas} \, \rm{yr}} \frac{d}{\rm{kpc}} km \ s^{-1},
\end{equation}
\noindent
where $v_{\ell, HC}$ is the transverse velocity relative to the Sun. To transform these velocities to a coordinate system in the Galactic center $v_{\ell, GC}$, we use the relation,

\begin{equation}
    \label{eqn:v_l}
    v^{*}_{\ell, GC} = v^{*}_{\ell, HC} - U_{\odot} \sin{\ell} \cos{b} + (V_{\odot} + V_{LSR}) \cos{\ell} \cos{b},
\end{equation}

\begin{figure*}
\includegraphics[width=\hsize]{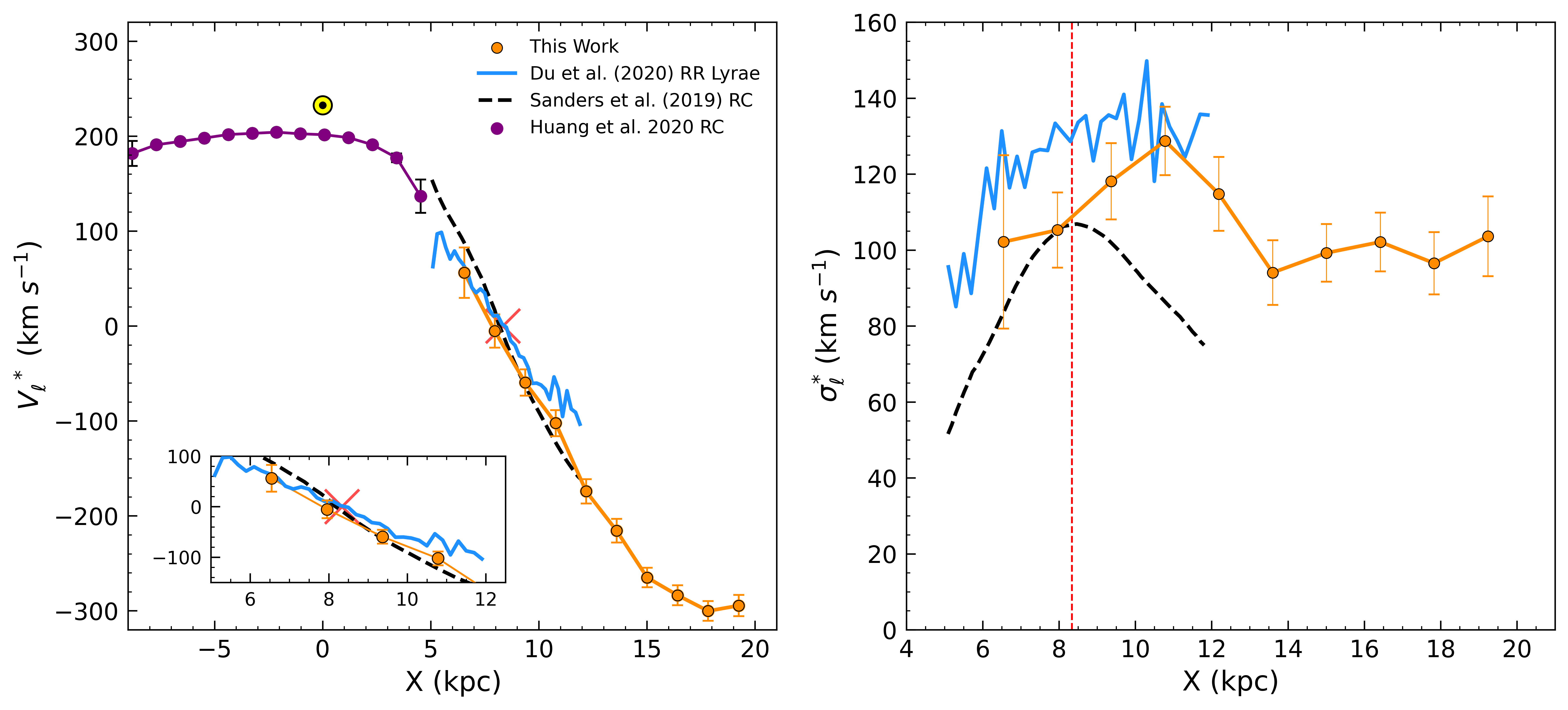}

    \caption{Rotation curve of the Galaxy as seen by our Mira sample. \textit{Left}: Galactocentric transversal velocity \textit{Right}: Velocity dispersion of the Mira sample. We add the corresponding curves from \citet{SandersRC_rot} (black) and \citet{Du+2020} (blue). We include the rotation curve derived from RC stars from \citet{Huang+2020_lamost} (purple). The velocity and position of the Sun are showcased in yellow. The red cross and the dashed line indicate the position of the Galactic center at a value of $V^{*}_{\ell} = 0$.} 
    \label{fig:rotation_curve}
\end{figure*}

\noindent
where $(U_{\odot}, V_{\odot}, W_{\odot})$ is the solar peculiar velocity relative to the Local Standard of Rest ($V_{LSR}$). Here, we adopted the values (11.1, 12.24,
7.25) $\rm{km} \ s^{-1}$ reported by \citet{Schonrich2010} for the solar peculiar velocity, and $V_{LSR} = 232.8 \ \rm{km} \ s^{-1}$ \citep{VLSR_mcmillan}. We bin the sample by distance, including only stars with P$<$400 days, errors in distance $\leq$10$\%$, and errors in $\mu^{*}_{\ell}$$\leq$1 mas/yr. These cuts result in 776 stars. Figure~\ref{fig:rotation_curve} shows their average transverse velocity (left), along with the velocity dispersion within each bin (right). We also include rotation curves based on RR Lyrae (\citealp{Du+2020}; average of metal-rich and metal-poor components) and Red Clump (RC) stars \citep{SandersRC_rot}. To compare with the rotation of the near side of the Galactic disk, we calculate the rotation velocities for the sample of RC stars presented in \citet{Huang+2020_lamost}. We only focus on stars that have ages comparable to our sample (i.e., age $\geq$ 5 Gyr) as the adopted cut in a period of 400 days translates into an age of $\sim$5.6 Gyr. We see a nice continuity between the rotation curves of nearby RC stars, bulge RR Lyrae, bulge RC stars, and Miras. 

Specifically, we find great agreement with the RR Lyrae-based rotation curve from \citep{Du+2020} for the Mira closer to the Galactic center. For further distances, at $\rm{X} \sim 11 \ \rm{kpc}$, we see that the shape of the rotation curve resembles the distribution of RC stars by \citet{SandersRC_rot}. At distances larger than $\sim 15 \ \rm{kpc}$, we find that the rotation curve flattens. Even though this effect is expected \citep{mroz+19}, this can also be attributed to the lower sensitivity of proper motions at large distances.

We showed in Fig.~\ref{fig:PL_comparison} that the distance distribution obtained with the MW-based P-L by \citet{SandersPL2023} would be similar to the present one. However, we note that for the rotation curve in Fig.~\ref{fig:rotation_curve} with the MW-based P-L, the derived distances would be larger, particularly for the Miras in the outer disk. This directly impacts the $V_{\ell}$ as shown in Eq. (\ref{eqn:v_l}). This, combined with the given PMs, results in larger rotation velocities ($>$300 km/s) at large R$_{\rm GC}$. In that case, the rotation curve in Fig.~\ref{fig:rotation_curve} would not be symmetric and would not show the expected flattening at V$^{*}_l$$\sim$$-$250 km/s. This is the main reason
why our data favor the LMC-based P-L relation, over the one based on Gaia parallaxes for nearby Miras.

\section{Discussion and conclusions}
\label{sec:Discussion_Conclusions}

We developed an algorithm to identify and characterize Mira candidates in the VVV survey, filtering candidates from all available Mira-like variables in order to match a validated sample of bona fide Miras.
We used the photometric information from both the VVV survey and WISE to obtain average magnitudes in multiple filters. After a final validation through visual inspection, we found 3602 Miras in the VVV survey that we consider as a pure sample.

We compiled and released, in this work, the largest catalog of variable stars to date at the far side of the Galactic Disk, providing mean magnitudes, distances, and PMs. 

The designed GP algorithm is tailored to find Mira candidates based on the \kssp band time series of the VVV survey. The modeled \kssp band is then used to determine average magnitudes in other VVV filters and WISE data. We also note that our choice of modeling for the long median trends in Miras was a simple polynomial to favor the detection of Miras with reliable average magnitudes. As a consequence, Miras with complex median magnitude trends might have been neglected from the analysis. 

We used the validated VarZoo sample, imposing a higher likelihood threshold to determine a Mira-based extinction law in the inner region of the Galaxy. We found great agreement with the extinction laws found by other studies such as \citet[classical cepheids]{Minniti+2020} and \citet[multiple variable stars]{Majaess+16}. 

Concerning the distinction between O-rich and C-rich Miras, we analyzed the separation in color excess planes and found clear differences between bona fide C-rich and O-rich Miras as shown in Fig. \ref{fig:OC_CC_separation}. After disentangling the populations, we derive a new extinction law and
compute distances to the O-rich Miras, finding most of the sample in the bulge region past the Galactic center, at the far side of the Galactic disk, as seen in Fig. \ref{fig:distances}. The kinematic information via PMs is consistent with the rotation of the Galaxy, as determined using several probes on this side of the disk (see Fig. \ref{fig:rotation_curve_pms}).

To exploit the correlation between the period and the age of the sample, we made bins by different ages and studied the distance distribution. As shown in Fig. \ref{fig:insideout}, we found a strong correlation between the distances and age of the sample, with younger stars lying at larger Galactocentric distances. This is consistent with the inside-out picture of the formation of the Galactic disk. We also investigated the warping of the Galactic disk as traced by the Mira sample in Fig. \ref{fig:warp}, finding that both the magnitude and the location of the warp are consistent with other studies based on younger tracers \citep{skowron19, Dekany+19, Minniti+2021}.

We studied the Galactic rotation curve by means of the transverse velocities, as shown in Fig. \ref{fig:rotation_curve}. We found great consistency in the bulge region with other tracers, especially with RR Lyrae \citep{Du+2020}. This consistency can be understood as the shorter period (older) Miras in our sample are more dominant closer to the Galactic center, tracing older populations in the same age bracket as RR Lyrae. At X$\sim$12 kpc, the average age of Miras, per bin, resembles that of RC stars ($\lesssim 9 \ \rm{Gyr}$), and we see a nice agreement with the rotation curve from \citet{SandersRC_rot}. For the furthest Mira (X$\sim$14 kpc) we see a flattening of the rotation curve at V$^{*}_{\ell}$$\sim$$-$260 km/s, consistent with the flattening of the rotation curve observed with Classical Cepheids \citep{mroz+19}. However, this could also be attributed to the sensitivity of PM-based velocities at these large distances. Still, the shape of the rotation curve is consistent with RC stars from \citet{Huang+2020_lamost} in the near side of the Galactic disk, showing a symmetric shape with respect to the Galactic center. We also observed that the velocity dispersion increases close to the Galactic center, then decreases as we approach the Galactic disk. The dispersion remains flat at higher distances due to the errors in PMs introducing a higher dispersion.

The catalog presented in this study represents the largest sample of variable stars at the far side of the Galactic disk and opens an opportunity to study intermediate-age and old stellar populations in this under-explored region of the MW.

\begin{acknowledgements}

We would like to thank the anonymous referee for suggestions that helped to improve this manuscript. We gratefully acknowledge the help from Massimo Griggio for the supplied data. R.A. acknowledges Manuel Cavieres, Francisco Jara-Ferreira, Nicolás Cristi and Benjamín Silva for the insightful discussions.

R.A. acknowledges support for this work by the National Agency for Research and Development (ANID) BASAL Center for Astrophysics and Associated Technologies (CATA) through grants AFB170002, ACE210002, and FB210003, and by the ANID Millenium Science Initiative, ICN12\_009 and AIM23-0001, awarded to the Millennium Institute of Astrophysics (MAS).
M.Z. and A.R.A. acknowledges support from DICYT through grant 062319RA and from ANID through FONDECYT Regular grant No. 1230731. J.O.C. acknowledges support from the ANID Doctorado Nacional grant 2021-21210865 and by ESO grant SSDF21/24. F.G. gratefully acknowledges support from the French National Research Agency (ANR)- funded projects ”MWDisc” (ANR-20-CE31-0004) and ”Pristine” (ANR-18-CE31-0017). A.V.N. acknowledges support from ANID Scholarship Program Doctorado Nacional 2020–21201226.

Based on observations collected at the European Southern Observatory under ESO programs 0103.D\-0386(A), 105.20MY.001, 179.B\-2002, and 198.B\-2004. We gratefully acknowledge the use of data from the VVV ESO Public Survey program ID 179.B-2002 taken with the VISTA telescope and data products from the Cambridge Astronomical Survey Unit (CASU). The VVV Survey data are made public at the ESO Archive. Based on observations taken within the ESO VISTA Public Survey VVV, Program ID 179.B-2002. This publication makes use of data products from the Wide-field Infrared Survey Explorer, which is a joint project of the University of California, Los Angeles, and the Jet Propulsion Laboratory/California Institute of Technology, and NEOWISE, which is a project of the Jet Propulsion Laboratory/California Institute of Technology. WISE and NEOWISE are funded by the National Aeronautics and Space Administration.

The Geryon cluster at the Centro de Astro-Ingenieria UC was extensively used for the calculations performed in this paper. BASAL CATA PFB-06, the Anillo ACT-86, FONDEQUIP AIC-57, and QUIMAL 130008 provided funding for several improvements to the Geryon cluster

This publication makes use of data products from the Two Micron All Sky Survey (2MASS), which is a joint project of the University of Massachusetts and the Infrared Processing and Analysis Center/California Institute of Technology, funded by the National Aeronautics and Space Administration and the National Science Foundation. 

This work has also made use of data from the European Space Agency (ESA) mission
{\it Gaia} (\url{https://www.cosmos.esa.int/gaia}), processed by the {\it Gaia}
Data Processing and Analysis Consortium (DPAC,
\url{https://www.cosmos.esa.int/web/gaia/dpac/consortium}). Funding for the DPAC
has been provided by national institutions, in particular the institutions
participating in the {\it Gaia} Multilateral Agreement.

It also made use of NASA’s Astrophysics Data System and of the VizieR catalog access tool, CDS, Strasbourg, France \citep{simbad}.  The original description of the VizieR service was published in \citep{vizier}. 
Finally, we acknowledge the use of the following publicly available software: Celerite \citep{Celerite}, TOPCAT \citep{topcat}, pandas \citep{pandas}, IPython \citep{ipython}, numpy \citep{numpy}, matplotlib \citep{matplotlib}, Astropy, a community developed core Python package for Astronomy \citep{astropy1,astropy2}, and Aladin sky atlas \citep{aladin1, aladin2}. 

\end{acknowledgements}


\bibliographystyle{aa}
\bibliography{mybiblio} 

\begin{appendix}
\section{Gaussian Process Modeling.}
    
Previous studies have found great success using GPs to model light curves of several types of stellar variability \citep{GP_asterosism_data}, including Mira variables \citep[S22]{He2016, Yuan+17}. A general GP can be described as a stochastic model consisting of a mean function ($\mu_{\theta}(x)$) and a covariance generally named Kernel ($\mathcal{K}(\alpha, \theta)$). The set of parameters from the mean function and the Kernel functions ($x, \theta, \alpha$) can be optimized, maximizing the likelihood function ($\ln{\mathcal{L}}(\theta, \alpha)$).

GP algorithms are generally avoided for large dataset applications due to their computational complexity. However, limiting the problem to one-dimensional data and using combinations of stationary exponential kernels greatly improves the performance \citep{Celerite}. This approach is implemented in the python package \textsc{CELERITE} \citep{Celerite} chosen for this work. 

Each light curve that meets the variability criteria detailed in Sect. \ref{sec:Identification} is then fed to a grid of GPs with kernels described by Eq. (\ref{eqn:kernel}).

\begin{equation}
\label{eqn:kernel}
    \mathcal{K}_{ij} = \mathcal{K}_{\rm{SHO}} \sum_{i = 0}^{1} \mathcal{K}_{\rm{DRW}}^{i} + \sum_{j = 0}^{1} \mathcal{K}_{\rm{DRW}}^{j} + \mathcal{K}_{\sigma}
\end{equation}
\noindent
Where $\mathcal{K}_{SHO}$ is a damped harmonic oscillator kernel, $\mathcal{K}_{DRW}$ is a damped random walk kernel, and $K_{\sigma}$ is a white noise kernel. both $i$ and $j$ have allowed values of 0 and 1, meaning 4 possible kernel combinations were considered in the analysis. We select a kernel combination based on the AIC, thus rewarding the use of simpler models for the light curves. 

For each iteration of the model, the fitted period was recovered from the maximum value of the power spectral density (PSD) of the fitted kernel. Since all considered kernels only include 1 term corresponding to a periodic kernel, the peak of the PSD corresponds to the dominant frequency according to the model, from which the period is recovered.

We emphasize that we chose to only include one periodic kernel ($\mathcal{K}_{\rm{SHO}}$), so models will only fit for one strongly periodic signal, while accommodating for stochastic effects and noise. However, complex median trends than cannot be captured by the chosen kernel combination will report a low $\mathcal{L}$, and thus, were automatically neglected. As discussed in Sect. \ref{sec:Identification}, this is made by design to isolate distance tracers, at the expense of possible real Miras not being considered in the final catalog.

\section{Saturated and blended sources.}
\label{sec:saturation_blends}

\begin{figure}
\includegraphics[width=\hsize]{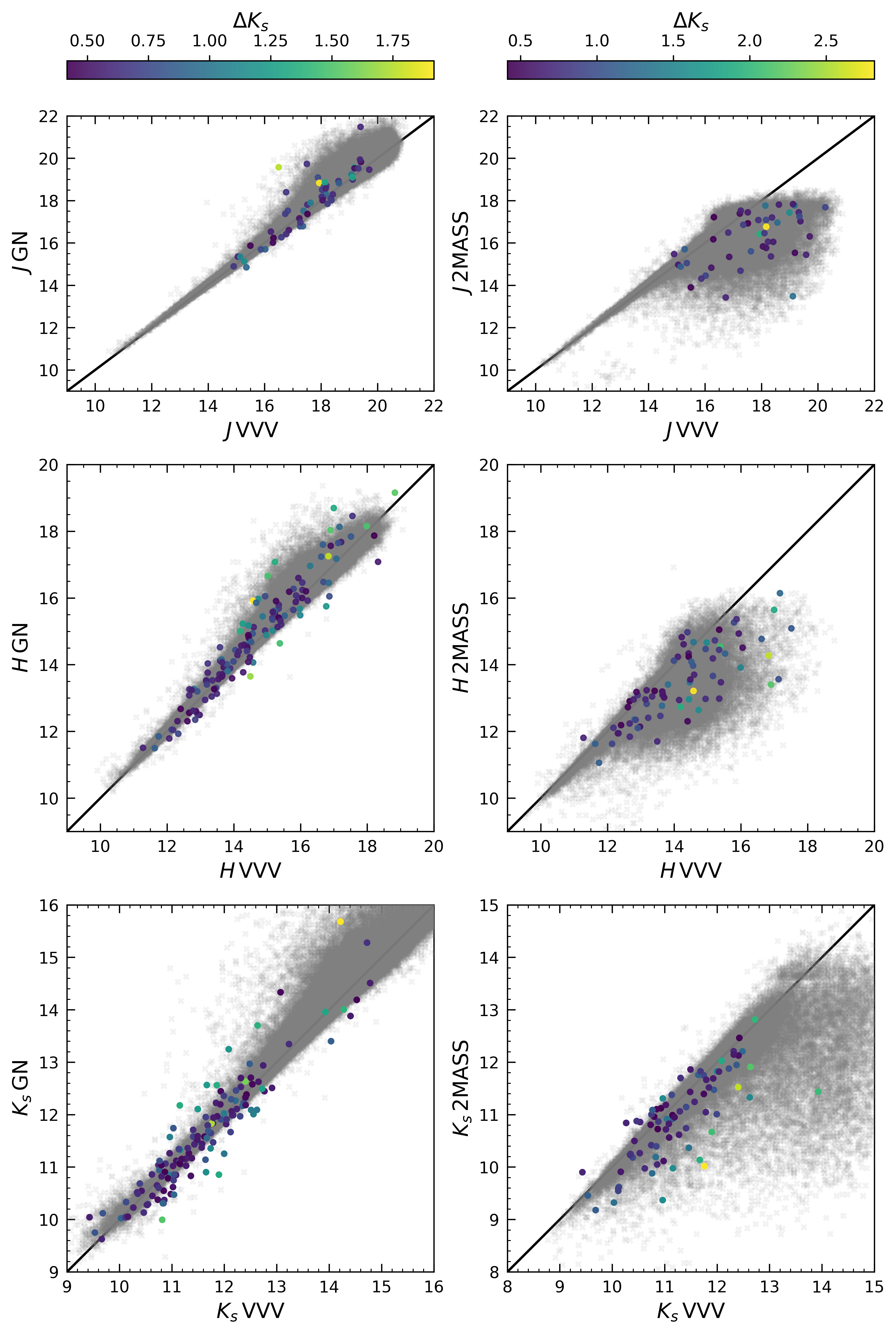}
    \caption{Comparison of magnitudes in GALACTICNUCLEUS, 2MASS, and VVV. Panels show the comparison of the same bands in the respective surveys. Gray points show all matched sources, and stars in our catalog present in both surveys are shown color-coded by our modeled amplitude in the \ksnosp band. Black solid lines indicate a 1:1 relation between magnitudes. We see no evidence of biases in the reported magnitudes between the surveys.}
    \label{fig:saturation}
\end{figure}

As previously noted, effects from saturation in the VVV survey begin to manifest for sources brighter than \ksnosp $\leq 12$ \citep{contrerasramos17}. To investigate this effect, we perform a crossmatch between our VVV catalog and GALACTICNUCLEUS \citep{GALACTICNUCLEUS}. We show the result of the crossmatch for non-variable sources in gray and for stars in our sample in Fig. \ref{fig:saturation}. Only VVV magnitudes have been corrected by the pulsation of the star, and thus, this introduces a scatter in the linear relation.

As mentioned, both VVV and GALACTICNUCLEUS begin to show saturation effects at \ksnosp $\leq 12$ and \ksnosp $\leq 11$, respectively. The shorter exposure times and smaller pixels in the HAWK-I camera from GALACTICNUCLEUS make saturation effects less apparent in this saturated regime. 

To further inspect the possible effect of saturation in the VVV photometry, we perform a match between the catalog and the 2MASS catalog \citep{2mass}. 2MASS begins to show saturation effects at \kssp $\leq 8.5$, effectively being saturation free at the \kssp magnitude range important for this study. The comparison between these catalogs is shown in the right hand side panels in Fig. \ref{fig:saturation}. We observe that the non-variable sources follow a linear relation in both catalogs, however, the confusion effects in the 2MASS catalog adds a significant scatter to the relation by over-estimating the brightness of the sources. Despite this, Miras in common between the catalogs do not show a clear bias or trend for stars fainter than \kssp $\leq 10$, and scatter around a 1:1 relation is driven by the lack of correction by pulsation in the 2MASS magnitudes and the confusion of the photometry. 

We conclude that even at the saturation regime of VVV, no significant bias in the Mira sources for magnitudes \kssp $\geq 10$ is observed. Thus, we deem the VVV magnitudes as reliable even at these saturated magnitudes for the purposes of this study. 

\section{WISE distances.}
\label{sec:WISE_distances}

\begin{figure*}
\includegraphics[width=\hsize]{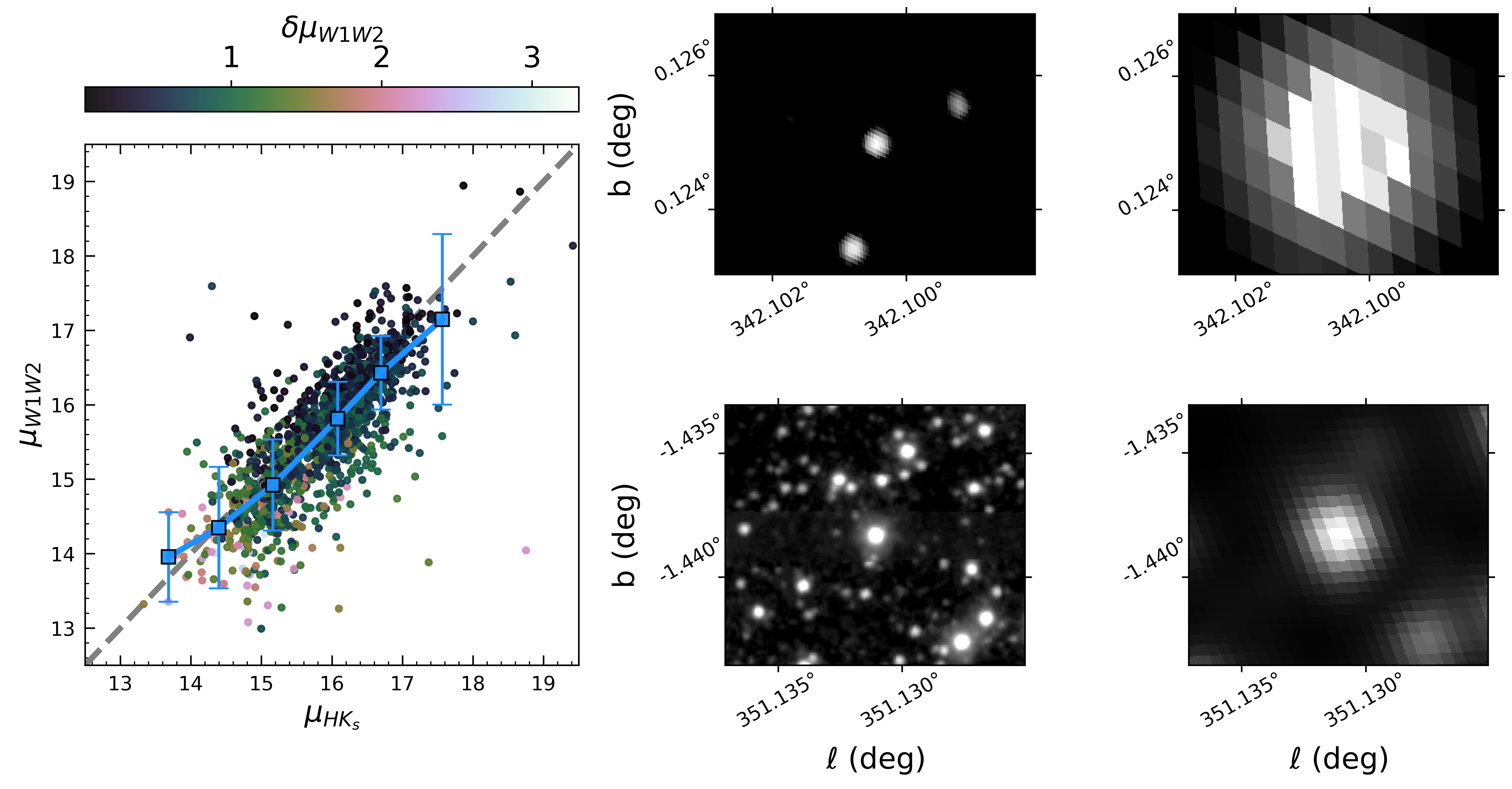}
    \caption{Distance modulus comparison between VVV based distances adopted here and WISE-based ones. \textit{Left}: Distance modulus comparison for Miras with WISE photometry available. Miras are shown in dots coloured by the error in the WISE based distance modulus. The gray dashed line shows a 1:1 relation between modulus. Blue squares show the median per bin. \textit{Middle panels}: VVV images in the H band of two Miras in our catalog. \textit{Right panels}: WISE images in the W1 band of the same Miras as the middle panels. Here, the blending caused by the larger PSF from WISE is clear.}
    \label{fig:WISE_comparison}
\end{figure*}

In this appendix, we discuss the distance determination for our sample using photometry from the WISE survey. We follow the same procedure described in Sect. \ref{sec:distances}, utilizing the average magnitudes in the W1 and W2 bands as detailed in Sect. \ref{sec:Multiband}, and applying the extinction law from \citet{WangExtLaw19}.

In the left panel of Fig.~\ref{fig:WISE_comparison}, we present the comparison between our adopted VVV-based distance modulus and the WISE-based ones. We exclude sources with $\mathrm{W1} < 8.1$ and $\mathrm{W2} \leq 6.7$, as these are heavily saturated in the respective bands \citep{WISEcorr}, leading to unreliable distance determinations. Overall, we observe good agreement between the adopted distances and those derived from WISE photometry; however, the VVV-based distances are systematically slightly larger across the entire range.

To investigate this discrepancy, we examine images from both the WISE survey (top panels in Fig.\ref{fig:WISE_comparison}) and the VVV survey (bottom panels in Fig.\ref{fig:WISE_comparison}). Due to the lower spatial resolution of WISE, nearby sources are blended with the target Mira variable, contributing additional flux to the observed magnitudes in the WISE bands. This blending effect results in brighter observed magnitudes, which leads to smaller distance modulus when compared to those derived from the higher-resolution VVV data. The additional flux from contaminant sources impacts both the observed magnitude and the derived extinction, shifting the distance distribution to lower values.

As mentioned for the VVV distances in Sect. \ref{sec:distances}, the majority of the error budget from Eq. (\ref{eqn:mod_err}) is dominated by the extinction term. For the WISE-based distances, the comparatively larger uncertainties in the extinction law from \citet{WangExtLaw19}, relative to our study, result in overall larger distance errors. Another interesting feature in Fig.~\ref{fig:WISE_comparison} is the larger errors in WISE distances for closer Miras. We attribute this behavior to the closer Miras being near the WISE saturation range, resulting in higher errors.

From this, we conclude that although WISE-based distances are generally reliable, blending near the Galactic plane has a considerable effect on the distances. The higher spatial resolution of the VVV survey reduces the impact of blending from nearby sources, resulting in more accurate distances.

\section{Catalog Example.}
\label{sec:catalog}

In this appendix, we include the header of  our final catalog, available at the CDS.

\begin{sidewaystable*}
  \flushleft
  \resizebox{ 0.97 \textheight}{!}{%
    {\small 
          \begin{tabular}{lrrllllllllllllllllllll}
    \toprule
    \hline
    ID & $\ell$ & b & P & J & $\delta \rm{J}$ & H & $\delta \rm{H}$ & \kssp & $\delta \rm{K_{s}}$ & W1 & $\delta \rm{W1}$ & W2 & $\delta \rm{W2}$ & W3 & $\delta \rm{W3}$ & W4 & $\delta \rm{W4}$ & d & $\delta \rm{d}$ & type & $\mu_{\ell}$ & $\mu_{b}$ \\

      & deg & deg & days & mag & mag & mag & mag & mag & mag & mag & mag & mag & mag & mag & mag & mag & mag & kpc & kpc &   & $\rm{mas yr}^-1$ & $\rm{mas yr}^-1$ \\
    \midrule
    \hline
    b33220745043 & -2.291557 & 0.011355 & 182.0 & 17.16 & 0.04 & 13.32 & 0.02 & 11.39 & 0.003 & - & - & - & - & - & - & - & - & 16.9 & 0.4 & O & -6.20 & -0.39 \\
    b33550389427 & 2.030869 & -0.078912 & 277.8 & 16.84 & 0.11 & 12.86 & 0.08 & 10.69 & 0.004 & 9.03 & 0.12 & 8.11 & 0.13 & - & - & - & - & 15.2 & 0.9 & O & -8.35 & -0.25 \\
    b32810812649 & -8.149293 & -0.309705 & 202.6 & 18.27 & 0.03 & 13.68 & 0.01 & 11.28 & 0.003 & 9.72 & 0.01 & 9.07 & 0.02 & 8.42 & 0.03 & 6.73 & 0.18 & 13.2 & 0.3 & O & -5.23 & 0.39 \\
    b30521329762 & -0.139154 & -1.590120 & 254.1 & 11.96 & 0.02 & 11.03 & 0.02 & 10.04 & 0.002 & - & - & - & - & - & - & - & - & 21.4 & 0.6 & C & -3.02 & -1.62 \\
    d0734062863 & -15.069632 & -0.240343 & 518.1 & - & - & 18.35 & 0.09 & 13.55 & 0.005 & 9.67 & 0.06 & 7.44 & 0.06 & 5.62 & 0.10 & 4.35 & 0.11 & 22.7 & 2.7 & C & -7.99 & 0.50 \\
    b30350392048 & -3.855943 & -2.254720 & 530.5 & - & - & 13.42 & 0.07 & 10.62 & 0.009 & 5.90 & 0.01 & 5.21 & 0.01 & 3.00 & 0.14 & 1.34 & 0.16 & 20.5 & 2.4 & C & -5.55 & -2.32 \\
    b33651156387 & 4.192921 & -0.012697 & 277.0 & 19.12 & 0.04 & 14.23 & 0.02 & 11.64 & 0.002 & 10.26 & 0.01 & 9.33 & 0.01 & 8.96 & 0.25 & 6.60 & 0.19 & 18.2 & 0.7 & O & -5.75 & -0.82 \\
    d11031023247 & -16.263470 & 0.798953 & 189.0 & 14.33 & 0.06 & 12.60 & 0.06 & 10.87 & 0.004 & 9.43 & 0.11 & 8.89 & 0.12 & 7.65 & 0.10 & 6.84 & 0.13 & 15.5 & 0.7 & O & -11.35 & -2.62 \\
    b25411639020 & 7.417047 & -6.940512 & 468.4 & 14.18 & 0.16 & 13.01 & 0.13 & 11.87 & 0.004 & 9.22 & 0.05 & 6.95 & 0.05 & 3.66 & 0.10 & 1.46 & 0.11 & 84.0 & 9.6 & C & -7.63 & 0.38 \\
    b33910291605 & 7.610559 & 0.114559 & 335.0 & 18.91 & 0.04 & 13.78 & 0.02 & 11.12 & 0.003 & 9.43 & 0.02 & 8.53 & 0.02 & 9.88 & 0.11 & - & - & 16.1 & 0.7 & O & -6.79 & -0.39 \\
    b31720969793 & -3.381637 & -0.574201 & 445.4 & 15.85 & 0.06 & 12.85 & 0.06 & 10.75 & 0.004 & 7.91 & 0.01 & 6.65 & 0.01 & 5.51 & 0.10 & 4.65 & 0.11 & 26.1 & 2.1 & C & -6.20 & -0.87 \\
    b30330874013 & -3.683423 & -2.413739 & 673.4 & - & - & 13.16 & 0.03 & 10.36 & 0.006 & 6.58 & 0.03 & 5.12 & 0.01 & 2.78 & 0.11 & 0.72 & 0.12 & 25.6 & 4.1 & C & -11.09 & 1.90 \\
    b33121184475 & -3.355265 & -0.060248 & 382.4 & 18.27 & 0.07 & 13.37 & 0.05 & 10.82 & 0.002 & 9.10 & 0.00 & 8.00 & 0.00 & 7.44 & 0.05 & 6.97 & 0.08 & 16.9 & 0.9 & O & -7.17 & -1.09 \\
    b31630239842 & -5.459128 & -0.701698 & 495.1 & - & - & 11.83 & 0.01 & 9.80 & 0.004 & 6.91 & 0.03 & 5.86 & 0.03 & 5.03 & 0.08 & 4.30 & 0.10 & 20.3 & 1.9 & C & -9.90 & -0.69 \\
    b31850962773 & -1.697197 & -0.581808 & 486.6 & 18.37 & 0.02 & 14.36 & 0.05 & 11.28 & 0.005 & 7.80 & 0.05 & 5.92 & 0.03 & 4.27 & 0.13 & 2.33 & 0.15 & 20.9 & 2.0 & C & -12.30 & 1.86 \\
    d10810440571 & -19.915534 & 0.028824 & 270.6 & 18.83 & 0.08 & 14.30 & 0.06 & 11.95 & 0.003 & - & - & - & - & - & - & - & - & 23.9 & 1.2 & O & -6.87 & 0.12 \\
    b33560444936 & 1.983577 & -0.358159 & 135.0 & 16.80 & 0.11 & 13.63 & 0.10 & 11.93 & 0.005 & 10.14 & 0.06 & 9.76 & 0.07 & - & - & - & - & 19.3 & 1.2 & O & -6.29 & 0.03 \\
    b33420338839 & 0.225009 & 0.031585 & 120.1 & 16.27 & 0.03 & 12.54 & 0.01 & 10.81 & 0.002 & 9.02 & 0.06 & 8.40 & 0.06 & - & - & - & - & 10.2 & 0.2 & O & -6.84 & -0.55 \\
    d1125039047 & -13.998910 & 0.470195 & 145.6 & 15.26 & 0.01 & 12.30 & 0.01 & 10.76 & 0.004 & 9.75 & 0.02 & 9.16 & 0.02 & 9.62 & 0.06 & - & - & 13.4 & 0.2 & O & -8.56 & 0.13 \\
    b34650374368 & -2.427427 & 1.048136 & 602.4 & - & - & 17.27 & 0.27 & 13.23 & 0.008 & 6.98 & 0.21 & 4.79 & 0.27 & 4.25 & 0.70 & 2.54 & 0.79 & 38.9 & 8.3 & C & -4.04 & -7.43 \\
    b3204024990 & 0.441987 & -0.635423 & 331.1 & - & - & 16.02 & 0.02 & 11.76 & 0.003 & 9.38 & 0.01 & 7.87 & 0.01 & 6.32 & 0.02 & - & - & 8.1 & 0.4 & O & -7.59 & 1.12 \\
    b32840438006 & -8.178010 & -0.176617 & 545.0 & 20.32 & 0.09 & 14.36 & 0.01 & 11.21 & 0.003 & 8.62 & 0.01 & 7.20 & 0.01 & 5.39 & 0.10 & 1.63 & 0.12 & 22.5 & 2.6 & O & -8.16 & -0.63 \\
    b33430475918 & 0.251130 & -0.230975 & 394.5 & 17.81 & 0.21 & 12.81 & 0.18 & 10.20 & 0.006 & 7.95 & 0.01 & 6.41 & 0.04 & 5.13 & 0.06 & 4.70 & 0.08 & 12.5 & 1.5 & O & 0.16 & 1.18 \\
    b38761431710 & -2.709962 & 4.586580 & 530.5 & 17.09 & 0.01 & 16.27 & 0.02 & 15.54 & 0.003 & - & - & - & - & - & - & - & - & 694.4 & 74.2 & C & -24.35 & 5.01 \\
    b32950343869 & -6.781690 & -0.001258 & 480.8 & - & - & 15.29 & 0.06 & 11.71 & 0.003 & 8.74 & 0.08 & 7.20 & 0.09 & 6.37 & 0.08 & 5.83 & 0.13 & 18.3 & 1.8 & O & -6.07 & -0.88 \\
    b33430439937 & 0.337004 & -0.157376 & 186.6 & 15.73 & 0.00 & 11.96 & 0.01 & 10.25 & 0.004 & 8.94 & 0.06 & 8.12 & 0.06 & - & - & - & - & 11.7 & 0.2 & O & -7.31 & 6.45 \\
    d11111216378 & -14.851840 & 0.079387 & 485.4 & 19.19 & 0.11 & 13.74 & 0.09 & 10.89 & 0.004 & 8.86 & 0.05 & 7.65 & 0.05 & 6.83 & 0.04 & 5.88 & 0.06 & 19.8 & 2.1 & O & -4.57 & 1.62 \\
    b32961614518 & -5.771306 & -0.290818 & 662.3 & 15.95 & 0.01 & 14.56 & 0.12 & 12.14 & 0.005 & 7.87 & 0.13 & 5.97 & 0.14 & 3.42 & 0.13 & 1.53 & 0.15 & 71.4 & 12.2 & C & -3.06 & -8.41 \\
    b32930449905 & -6.902989 & -0.210528 & 320.0 & 17.26 & 0.04 & 12.90 & 0.04 & 10.58 & 0.002 & 8.97 & 0.00 & 8.11 & 0.00 & 7.77 & 0.03 & 6.81 & 0.09 & 14.9 & 0.7 & O & -8.83 & -0.13 \\
    d0752138561 & -11.607840 & -0.065624 & 175.0 & 13.05 & 0.06 & 11.14 & 0.05 & 9.98 & 0.003 & 9.14 & 0.01 & 8.77 & 0.01 & 8.37 & 0.02 & - & - & 13.6 & 0.5 & O & -8.55 & -0.43 \\
    b33241627731 & -1.237215 & -0.134515 & 490.2 & 17.68 & 0.11 & 13.97 & 0.09 & 10.97 & 0.003 & 10.72 & 0.05 & 8.48 & 0.05 & 7.69 & 0.03 & 3.78 & 0.02 & 19.1 & 2.0 & O & -5.54 & -0.15 \\
    b33621531138 & 4.417195 & 0.043633 & 823.1 & - & - & 16.02 & 0.04 & 11.69 & 0.008 & 5.92 & 0.24 & 3.70 & 0.26 & 0.37 & 0.43 & -1.59 & 0.49 & 25.0 & 5.2 & C & -3.13 & 0.48 \\
    b34411683989 & -4.361689 & 0.654729 & 475.1 & - & - & 18.01 & 0.10 & 12.50 & 0.004 & 8.97 & 0.06 & 6.74 & 0.07 & 4.96 & 0.03 & 3.00 & 0.04 & 8.0 & 0.9 & O & -11.89 & 3.06 \\
    \hline
    \end{tabular}
    }
  }
  \caption{Median magnitudes and Proper Motions for Miras in our catalog.}
  \label{tab:rotated_scaled}
\end{sidewaystable*}

\end{appendix}

\label{lastpage}
\end{document}